\documentclass[twocolumn,epjc3]{svjour3}  
\smartqed  
\RequirePackage{graphicx}
\journalname{Eur. Phys. J. A}
\usepackage{color}
\usepackage{amsmath,amssymb}
\usepackage[dvipsnames]{xcolor}
\usepackage{microtype}
\usepackage{tikz}

\usepackage{youngtab}
\usepackage{young}
\usepackage{ytableau}
\usepackage{bm}
\usepackage[normalem]{ulem}
\usepackage[colorlinks=true,allcolors=Blue,pdfusetitle]{hyperref}
\usepackage{braket}
\usepackage{cite}

\usepackage{tabularx}
\usepackage{slashbox}

\usepackage[utf8]{inputenc}
\usepackage{fourier} 
\usepackage{array}
\usepackage{makecell}

\begin{document}

\title{Quantum information and quantum simulation of neutrino physics}


\author{
A. B. Balantekin\thanksref{e1,addr1}
\and
Michael J. Cervia\thanksref{e2,addr2,addr3} 
\and 
Amol V. Patwardhan\thanksref{e3,addr4}
\and
Ermal Rrapaj\thanksref{e4,addr5,addr6,addr7} 
\and
Pooja Siwach\thanksref{e5,addr1} 
}

\thankstext{e1}{e-mail: baha@physics.wisc.edu}
\thankstext{e2}{e-mail: cervia@gwu.edu}
\thankstext{e3}{e-mail: apatward@slac.stanford.edu}
\thankstext{e4}{e-mail: ermalrrapaj@berkeley.edu}
\thankstext{e5}{e-mail: psiwach@physics.wisc.edu}


\institute{
University of Wisconsin, 1150 University Ave, Madison, WI 53706 \label{addr1}
\and
George Washington University, 725 21st St NW, Washington, DC 20052 \label{addr2}
\and
University of Maryland, College Park, MD, USA 20742 \label{addr3}
\and
SLAC National Accelerator Laboratory, 2575 Sand Hill Rd, Menlo Park, CA 94025 \label{addr4}
\and       
ERSC, Lawrence Berkeley National Laboratory, Berkeley, California 94720, USA\label{addr5}
\and
RIKEN iTHEMS, Wako, Saitama 351-0198, Japan\label{addr6}
\and
University of California, Berkeley, CA 94720-7300 \label{addr7}
}

\date{Received: date / Accepted: date}

\maketitle

\abstract{
In extreme astrophysical environments such as core-collapse supernovae and binary neutron star mergers, neutrinos play a major role in driving various dynamical and microphysical phenomena, such as baryonic matter outflows, the synthesis of heavy elements, and the supernova explosion mechanism itself. The interactions of neutrinos with matter in these environments are flavor-specific, which makes it of paramount importance to understand the flavor evolution of neutrinos. Flavor evolution in these environments can be a highly nontrivial problem thanks to a multitude of collective effects in flavor space, arising due to neutrino-neutrino ($\nu$-$\nu$) interactions in regions with high neutrino densities. A neutrino ensemble undergoing flavor oscillations under the influence of significant $\nu$-$\nu$ interactions is somewhat analogous to a system of coupled spins with long-range interactions among themselves and with an external field (\lq long-range\rq\ in momentum-space in the case of neutrinos). As a result, it becomes pertinent to consider whether these interactions can give rise to significant quantum correlations among the interacting neutrinos, and whether these correlations have any consequences for the flavor evolution of the ensemble. In particular, one may seek to utilize concepts and tools from quantum information science and quantum computing to deepen our understanding of these phenomena. In this article, we attempt to summarize recent work in this field.
}
{ Furthermore, we also present some new results in a three-flavor setting, considering complex initial states.}
\keywords{Neutrino oscillations \and Quantum computing \and Many-body systems \and Symmetries}
%
\noindent
%


\section{Neutrinos in extreme astrophysical environments} \label{sec:intro}

{ Neutrinos, owing to their feeble interactions with matter, are highly efficient at transporting energy, entropy, and lepton number in extreme astrophysical settings such as core-collapse supernovae and binary compact object mergers (binary neutron star or black hole - neutron star), as well as during certain epochs in the early universe (e.g., see Refs.~\cite{Janka:2006fh,Burrows:2020qrp,Fuller:2022nbn,Foucart:2022bth,Kyutoku:2017voj,Grohs:2015tfy}). As a result, they are expected to play a key role in influencing the dynamics and nucleosynthesis in these environments.

The flavor-specific interactions that govern the neutrino transport and determine the free neutron and proton abundances in these environments are charged current (anti-)neutrino captures that convert neutrons into protons, and vice versa.
}
{Given the typical temperatures and densities of these environments, neutrinos decouple with energies of $\mathcal{O}(1\text{--}10)\,$MeV, and therefore, the charged-current interactions of $\mu$ and $\tau$ flavor (anti-)neutrinos are 
energetically suppressed. 
Since these charged current processes govern the energy transport as well as the neutron-to-proton ratio (or equivalently electron fraction), which in turn determines nucleosynthesis yields (e.g., ~\cite{Surman:2003qt,Martinez-Pinedo:2017ksl,Kajino:2014bra,Frohlich:2015spx,Langanke:2019ggn,Roberts:2016igt,Steigman:2012ve,Grohs:2015tfy}), the flavor-asymmetric nature of charged-current capture implies that a complete understanding of neutrino flavor evolution in these environments is crucial 
\cite{Qian:1993dg,Yoshida:2006qz,Duan:2010af,Kajino:2012zz,Wu:2014kaa,Sasaki:2017jry,Balantekin:2017bau,Xiong:2019nvw,Xiong:2020ntn,George:2020veu}.

In this paper we first summarize recent progress in our understanding of 
neutrino oscillations in extreme astrophysical environments, particularly the quantum many-body aspect of collective neutrino oscillation physics driven by $\nu$-$\nu$ interactions in high neutrino densities (Secs.~\ref{sec:coll}--\ref{sec:qsim}). In Sec.~\ref{sec:3fla}, we show some new results of many-body neutrino calculations in a three-flavor setup. Finally, we offer concluding remarks in Sec.~\ref{sec:concl}. 
}


\section{Collective neutrino oscillations} \label{sec:coll}

{ Neutrinos experience flavor oscillations thanks to a misalignment between the propagation (mass) eigenstates and the eigenstates of the weak interaction (flavor). This misalignment, present even in vacuum, can be modified in interesting ways when neutrinos pass through a medium. Where the medium consists of baryons and charged leptons, neutrino coherent forward-scattering with these particles can lead to suppressed or resonantly enhanced flavor conversion through the Mikheyev-Smirnov-Wolfenstein (MSW) mechanism~\cite{Wolfenstein:1977ue,Mikheyev:1985zog,Mikheev:1986if}. This mechanism is the widely accepted solution for explaining the observed solar neutrino deficit. Even more fascinating is the scenario where the neutrinos themselves form a significant component of the background, as can be the case in extreme astrophysical environments like core-collapse supernovae, binary compact object mergers, and the early universe. In this situation, neutrinos experience an additional forward-scattering potential on account of pairwise weak neutral current interactions with other neutrinos~\cite{Notzold:1987ik,Fuller:!987aa}. In the low-energy limit, this can be represented by the four-Fermi effective operator
\begin{equation} \label{eq:nunuint}
    H_\mathrm{int} \equiv \frac{G_F}{\sqrt{2}} \sum_{\alpha,\beta} (\overline{\nu}_\alpha \gamma^\mu\nu_\alpha) (\overline{\nu}_\beta\gamma_\mu\nu_\beta),
\end{equation}
Unlike the interactions with baryons and charged leptons, which are diagonal in the flavor basis, the coherent $\nu$-$\nu$ scattering can give rise to both diagonal and off-diagonal potentials, each of which depend on the flavor composition of the neutrino background~\cite{Pantaleone:1992xh,Pantaleone:1992eq,Samuel:1993}. In this sense, the background itself becomes dynamical and the flavor evolution histories of the interacting neutrinos thus become coupled to one another. This coupling introduces a non-linearity and a nontrivial geometrical complexity to the flavor evolution problem, giving rise to many kinds of collective oscillation modes in flavor space. See, for example, the reviews in Refs.~\cite{Duan:2009cd,Duan:2010,Chakraborty:2016yeg,Tamborra:2020cul,Richers:2022zug} and references therein. In particular, a lot of the recent literature deals with the aptly named \lq\lq fast flavor transformations\rq\rq\, which occur as a result of nontrivial angular distributions in the neutrino flavor field (reviewed in Refs.~\cite{Chakraborty:2016yeg,Tamborra:2020cul,Richers:2022zug}). Flavor instabilities arising in this manner grow on a much faster timescale compared to the instabilities present in physical setups with a greater degree of angular symmetry, hence the moniker \lq\lq fast\rq\rq.

Another feature of this problem that has received attention in recent years is the quantum nature of these collective effects, as part of a broader focus on quantum simulations in high energy physics~\cite{Bauer:2022hpo}. The pairwise interactions between neutrinos that give rise to the coherent $\nu$-$\nu$ scattering potentials assume the form of a spin-spin coupling, when expressed using \lq\lq isospin\rq\rq\ operators defined in the Hilbert space of neutrino flavor states (described in detail in the following section). Depending on whether one considers two or three flavors, the isospin operators for each neutrino form a SU(2) or SU(3) algebra. The system of interacting neutrinos then constitutes a quantum many-body system, with the size of the associated Hilbert space scaling exponentially with the number of particles in the system ($2^N$ or $3^N$ for a system of $N$ neutrinos and anti-neutrinos in two or three flavors, respectively).

Owing to the immense computational difficulty of analyzing a many-body system with an appreciable number of neutrinos, it becomes necessary to invoke certain simplifying assumptions. One common approach involves postulating that the effect of multi-particle quantum correlations on the flavor evolution will remain negligible in the large-$N$ limit, as a result of a large number of random phases being added incoherently. With this assumption, one is allowed to construct a simplified Hamiltonian, where the two-particle interaction operator is replaced with an effective one-particle interaction with a \lq\lq mean-field\rq\rq\ representing all of the background particles~\cite{Samuel:1993,Sigl:1993ctk,Qian:1994wh}. In this \lq\lq Random phase approximation\rq\rq, the effective dimensionality of the state-variable space is reduced to $n_fN$ (where $n_f$ is the number of flavors), thus making the problem much more tractable numerically.

However, this simplification naturally raises the question as to whether such an assumption can lead to the exclusion of any crucial physics from the problem. Several recent studies have attempted to answer this question using a variety of different physical setups and numerical methods. In what follows, we review some of the recent progress.

}

\section{Physical setup and neutrino Hamiltonian}

{
 In much of the recent literature investigating many-body neutrino correlations, the neutrinos are modeled as interacting plane waves in a box of volume $V$ (which can be taken to be time-dependent to mimic the decrease of the neutrino number density with distance from the source). In this picture, the finite size of the neutrino wave-packet (and consequently, the finiteness of the interaction interval between any two neutrinos) is not taken into account. Therefore, this setup may not be fully suitable for analyzing certain features of this problem, such as effects of incoherent scattering on the neutrino flavor conversion~\cite{Shalgar:2023ooi}. Nevertheless, it can be a useful initial step for illustrating the feedback effect from non-trivial $\nu$-$\nu$ correlations on the flavor dynamics, in a simplified setting. In order to fully validate or invalidate the mean-field approximation, more detailed analyses including wave-packet effects may be needed in the future.
}

{The Hamiltonian describing a system of interacting neutrinos can be written in terms of generators of $SU(n_f)$ where $n_f$ are the number of neutrino flavors. 
{
 
For instance, in a two-flavor model, the SU(2) generators, in terms of the Fermionic creation and annihilation operators, are defined as~\cite{Balantekin_2006}
\begin{gather}
	{J}_{\mathbf{p}}^{+}= a_{1}^{\dagger}(\mathbf{p})\,a_{2}(\mathbf{p})~, \\ 
  	{J}_{\mathbf{p}}^{-}= a_{2}^{\dagger}(\mathbf{p})a_{1}(\mathbf{p})~, \\ 
	{J}_{\mathbf{p}}^z=\frac{1}{2}\left(a_{1}^{\dagger}\,(\mathbf{p})a_{1}(\mathbf{p})-a_{2}^{\dagger}\,(\mathbf{p})a_{2}(\mathbf{p})\right)~, 
    \label{eq:nfis} 
\end{gather} 
in the mass basis. Alternatively, one may construct a flavor basis SU(2) algebra, with $a_e$ and $a_x$ replacing $a_1$ and $a_2$ (where $x$ represents an appropriate superposition of $\mu$ and $\tau$ flavors). In the spin-$1/2$ representation, one can write these operators in terms of Pauli matrices: i.e., $\vec{J}_\mathbf{p} = \vec \sigma_\mathbf{p}/2$, where $\vec\sigma_\mathbf{p}$ is a vector of Pauli matrices defined in the subspace of the neutrino with momentum $\mathbf{p}$. In what follows, we also refer to these SU($n_f$) generators as neutrino \lq\lq isospin\rq\rq\ operators. In the two-flavor context, isospin \lq\lq up\rq\rq\ and \lq\lq down\rq\rq\ can refer to $\ket{\nu_1}$ and $\ket{\nu_2}$, respectively, in the mass basis, or $\ket{\nu_e}$ and $\ket{\nu_x}$ in the flavor basis.
In this two-flavor picture, a Hamiltonian consisting of terms that represent vacuum mixing as well as $\nu$-$\nu$ interactions\footnote{ Here, we exclude the term representing neutrino interactions with ordinary matter (e.g., baryons and charged leptons), since it has a structure that is conceptually similar to the vacuum oscillation term---i.e., consisting of individual neutrinos interacting with a background. In regimes where collective oscillation effects typically dominate, this matter-interaction term can be \lq\lq rotated away\rq\rq\ with a suitable change-of-basis transformation{, resulting in a modified mixing angle and mass-squared splitting compared to the corresponding values in vacuum}.} can be written as
\begin{equation} \label{eq:ham}
H = \sum_{\mathbf{p}}\omega_\mathbf{p} \, \vec{B}\cdot\vec{J}_{\mathbf{p}} + \frac{\sqrt{2}G_{F}}{V}\sum_{\mathbf{p},\mathbf{q}}\left(1-\mathbf{\widehat p}\cdot\mathbf{\widehat q}\right)\,\vec{J}_\mathbf{p}\cdot\vec{J}_\mathbf{q}~,
\end{equation}
where $\vec{B}$ can be interpreted as a \lq\lq background field\rq\rq\ that points along the direction of the mass basis. It is equal to $(0,0,-1)$ in the mass-basis representation, or $(\sin2\theta,0,-\cos2\theta)$ in the flavor-basis representation. $\omega_\mathbf{p} = {\delta m^2}/{(2|\mathbf{p}|)}$ are the vacuum oscillation frequencies for neutrinos with momenta $\mathbf{p}$, with $\delta m^2$ being the mass-squared difference between the eigenstates.  $\mathbf{\widehat p}$ and $\mathbf{\widehat q}$ are the unit vectors along the momenta of interacting neutrino pairs, and $V$ is the quantization volume. One can define a $\nu$-$\nu$ coupling parameter $\mu \equiv \sqrt{2} G_F N/V$, where $N$ is the total number of interacting neutrinos. 
}
The strength of the $\nu$-$\nu$ interactions thus depends on the neutrino number density and the intersection angle between their trajectories of the interacting neutrinos. This dependence introduces an additional geometric complexity to the problem, aside from the complexity associated with the exponential scaling of the Hilbert space. { Note that the Hamiltonian of Eq.~\eqref{eq:ham} consists only of terms that either preserve or exchange the momenta of interacting neutrino pairs (\textit{forward} and \textit{exchange} terms), since these can be added up coherently in the mean-field limit. Recent work has argued that these terms should not enjoy special status in an exact many-body calculation, unlike in the mean field limit, and, as a result, consideration of a more generalized Hamiltonian with interaction terms besides forward and exchange scattering is warranted~\cite{Johns:2023ewj}. This addition could be an important avenue to pursue in future studies.}
{

 In the mean field approach, the interaction term in the Hamiltonian is replaced with an effective one-particle operator, using the following prescription:
\begin{equation} \label{eq:meanfield}
    \vec{J}_\mathbf{p}\cdot\vec{J}_\mathbf{q} \approx \vec{J}_\mathbf{p}\cdot\langle \vec{J}_\mathbf{q} \rangle + \langle \vec{J}_\mathbf{p} \rangle \cdot\vec{J}_\mathbf{q} - \langle \vec{J}_\mathbf{p} \rangle \cdot \langle \vec{J}_\mathbf{q} \rangle.
\end{equation}
    This approximation in essence enforces that the wavefunctions of individual neutrinos remain uncorrelated through the course of the evolution { (assuming that the neutrino ensemble starts out in an uncorrelated state)}, and the system as a whole thereby remains in a direct product state, greatly reducing the effective number of independent amplitudes from $n_f^N$ to $n_fN$.
}

\section{Quantum dynamics of collective neutrino oscillations}

{
\subsection{\it Initial work}

It was recognized early on that $\nu$-$\nu$  interactions give rise to non-diagonal terms in the quantum many-body problem, which may not always be factorizable in terms of a one-particle effective approximation~\cite{Pantaleone:1992xh,Pantaleone:1992eq}. In subsequent attempts to ascertain the validity of the one-particle effective approximation~\cite{Bell:2003mg,Friedland:2003dv,Friedland:2003eh,Friedland:2006ke,McKellar:2009py}, the flavor evolution of interacting neutrinos was analyzed with two different approaches: (i) using two intersecting beams of neutrinos, where the flavor evolution was described in terms of a sequence of elementary scattering amplitudes, and (ii) using a neutrino ensemble represented as interacting plane waves in a box. 

Initial disagreement about the possibility of quantum entanglement developing among neutrinos~\cite{Bell:2003mg,Friedland:2003dv} culminated in the understanding that the timescales for the build-up of entanglement suggest the possibility of incoherent effects~\cite{Friedland:2003eh}. These conclusions were further generalized in Ref.~\cite{Friedland:2006ke}. On the other hand, these analyses employed several simplifications such as omission of the one-body terms in the Hamiltonian. Even in the mean-field approximation, the interplay between vacuum oscillations and $\nu$-$\nu$ interaction terms give rise to interesting collective phenomena such as \lq\lq spectral splits\rq\rq~\cite{Duan:2006jv,Duan:2006c,Duan:2007,Raffelt:2007,Raffelt:2007xt}. { Furthermore, it has been argued~\cite{McKellar:2009py} that even an incoherent timescale for the evolution does not necessarily preclude the presence of significant multi-particle correlations.} Therefore, the quantum many-body dynamics of collective neutrino oscillations
remains an interesting topic that merits further exploration.

These early results in the past predicted either a vanishingly small contribution~\cite{Friedland:2003eh,Friedland:2006ke} in the limit of large system size $N$, or substantial flavor evolution over time scales $\tau_F \sim \mu^{-1}\log(N)$ that can remain relevant for large systems~\cite{Bell:2003mg,Sawyer:2004}. More recently, the role of entanglement and quantum effects in the out-of-equilibrium dynamics~\cite{Eisert:2015} of neutrinos has received renewed interest (e.g., \cite{Cervia:2019,Rrapaj:2020,Roggero:2021asb} and subsequent works mentioned later in the text). Note that flavor oscillations on the time scale $\tau_F$ can be considered to be \lq\lq fast\rq\rq, analogous to the classification in the mean-field literature, wherein \lq\lq fast\rq\rq\ and \lq\lq slow\rq\rq\ oscillations refer to time scales $\sim\mu^{-1}$ and $\sim(\mu\omega)^{-1/2}$ (or $\omega$), respectively.
}

\subsection{\it Quantifying entanglement in a neutrino ensemble} \label{sec:quant}
{ 
To define the amount of entanglement in an interacting neutrino system, one must first devise a partition. For instance, one can consider all neutrinos travelling parallel to one another as a \lq\lq beam\rq\rq\ or \lq\lq sub-system\rq\rq and separate the system in terms of beams. 
Alternatively, one may simply separate each neutrino from the rest.
In either case, one can define the reduced density matrix of sub-system $A$ by taking a partial trace of the full density matrix $\rho$ over the complement\footnote{ For a multi-neutrino system in a pure quantum state denoted by wavefunction $\ket{\Psi}$, the density matrix of the entire system is  $\rho = \ket{\Psi}\bra{\Psi}$. The complement of sub-system $A$ is defined so that $A \cup A^c$ represents the full quantum system.}
of $A$: i.e., $\rho_A \equiv \mathrm{Tr}_{A^c}[\rho]$. For example, the reduced density matrix of a single neutrino of momentum $q$ can be given as 
\begin{equation}
    \rho_q= \sum\limits_{i_1,\ldots,\widehat{i_q},\ldots,i_{N}=1}^2\braket{\nu_{i_1}\ldots\widehat{\nu_{i_q}}\ldots\nu_{i_{N}}|\rho|\nu_{i_1}\ldots\widehat{\nu_{i_q}}\ldots\nu_{i_{N}}},
\end{equation}
where the $\,\,\widehat{}\,\,$ symbol denotes exclusion. The entropy of entanglement of sub-system $A$ with the rest of the ensemble is then defined as the von Neumann entropy of the reduced density matrix:
\begin{equation}
    S_A = -\mathrm{Tr}[\rho_A \log \rho_A] = -\sum_j \lambda^{(A)}_j \log \lambda^{(A)}_j,
\end{equation}
where $\lambda^{(A)}_j$ are the eigenvalues of the reduced density matrix $\rho_A$. 
Another important measure of entanglement is the R\'{e}nyi entropy, which is defined as
\begin{equation}
    \mathcal R_{\gamma, A} = \frac{1}{1-\gamma} \log [\mathrm{Tr}(\rho_A^\gamma)].
    \label{eq:renyi_entropy_gen}
\end{equation}
The von Neumann entropy can be expressed as R\'{e}nyi entropy in the limit of $\gamma\to 1$
\begin{equation}
     S_A = \lim_{\gamma\to 1} \mathcal R_{\gamma, A} = -\mathrm{Tr}[\rho_A \log (\rho_{A})].
    \label{eq:von_neumann_entropy_gen}
\end{equation}
In multi-beam systems, one can compactly represent the wavefunction of the system
in a flavor angular momentum basis, e.g., 
\begin{equation}
\ket{\Psi} = \sum_{m_A, m_B,\ldots} a_{m_A, m_B,\ldots} \ket{m_A, m_B,\ldots},    \label{eq:psi_angular_momentum_basis}
\end{equation}
where $m_A$ is the difference between the number of electron neutrinos versus the rest in beam $A$. As an example, the R\'{e}nyi entropy of beam $A$ is given as~\cite{Roggero:2022}, 
\begin{equation}
    \mathcal R_{\gamma, a}
    = \frac{1}{1-\gamma} \log \left[ \sum_{m_A} \left( \sum_{m_B,\ldots} |a_{m_A, m_B,\ldots}|^2 \right)^\gamma \right].
    \label{eq:R_beams}
\end{equation}

The evolution of the interacting neutrino system can be characterized in terms of the \lq\lq Polarization vectors\rq\rq of individual neutrinos, which are related to the expectation values of the neutrino isospin operators defined previously,\footnote{ In the case of neutrino beams, the definition of the isospin operators can be generalized to represent the entire beam: $\vec J_A = \sum_{q \in A} \vec J_q$. The polarization vector of the beam is then given by $\vec{P}_A = 2 \langle \vec{J}_A \rangle/N_A$.} i.e.,
$\vec{P}_q = 2 \langle \vec{J}_q \rangle$. In terms of these polarization vector components, the reduced density matrix for an individual neutrino can be written as
\begin{equation}
\rho_q = \frac{1}{2}
\begin{bmatrix}
1 + P_{q,z} & P_{q,x} - \mathrm{i}P_{q,y} \\
P_{q,x} + \mathrm{i}P_{q,y} & 1- P_{q,z}
\end{bmatrix},
\end{equation}
and therefore the entanglement entropy can be related to the length of the Polarization vector, in accordance with
\begin{equation}
    S(\omega_q)=-\frac{1-P_q}{2}\log\left(\frac{1-P_q}{2}\right)-\frac{1+P_q}{2}\log\left(\frac{1+P_q}{2}\right).
\end{equation}

Note that, in the mean-field limit, the reduced density matrix $\rho_q$ is simply a pure state of the neutrino $q$, i.e., $\rho_q^{(\text{MF})} = \ket{\psi_q}\bra{\psi_q}$, where $\ket{\psi_q}$ is the wavefunction of neutrino $q$. In that case, the density matrix $\rho_q$ is that of an unmixed state with $\mathrm{Tr}[\rho_q^2] = 1$, implying $P_q = 1$ and $S(\omega_q) = 0$. In the many-body case where neutrino $q$ may be entangled with the other neutrinos, $\rho_q$ takes the form of a mixed-state density matrix with $\mathrm{Tr}[\rho_q^2] < 1$ and $S(\omega_q) = 0$.
}

{ 
A closely related measure of entanglement comes in the form of left-right entanglement entropy, convenient particularly in the study of a two-beam model as noted first by Ref.~\cite{Roggero:2021asb}. 
When calculated via tensor network methods, this entanglement entropy can be expressed in terms of the singular values $v_k(b)$ ($k=1,\ldots,r_b$) encoding each bond $b$ between localized matrices of the { wave function} network {~\cite{Roggero:2021asb}}:
\begin{equation}
    S_{LR}(b) = -\sum_{k}v_k(b)^2\log\left[v_k(b)^2\right].
    \label{eq:left-right_entangle}
\end{equation}
One can show for matrix product states that $S_{LR}(b)\leq\log(r_b)$, where $r_b\leq n_f^{\min(b,N-b)}$ is the [truncated] dimension of bond $b$. 
{ For more detailed discussion of how to calculate these singular values and related entropy formulae, see e.g., Refs.~\cite{Nielsen:2011:QCQ:1972505,2011AnPhy.326...96S,PAECKEL2019167998,Vidal:2003lvx,Schuch:2008zza}. }

In kind, one can compute quantum negativity, 
\begin{equation}
    \mathcal{N}_q = \frac{1}{2}\left(\sqrt{\mathrm{Tr}\left[\left(\rho^{T_q}\right)^\dagger\rho^{T_q}\right]}-1\right),
\end{equation}
where $\rho^{T_q}$ is a partial transpose over the degrees of freedom for neutrino $q$. This entanglement measure was briefly considered in Ref.~\cite{Cervia:2019}, however, its behavior over time evolution was found to be qualitatively similar to entanglement entropy of each neutrino mode, while the latter was computationally much less expensive to find. 

Furthermore, one can consider measurements of entanglement to describe an entire global wave function, as opposed to the bipartite measures listed above. 
For example, the $n$-tangle~\cite{PhysRevA.63.044301} 
can be used to measure of entanglement globally in collective oscillations~\cite{Illa:2022zgu}. 
Functionally, one defines a $n$-tangle measure for a system of size $N$ as the overlap of a state and its complex conjugate with $n$ spin flips. 
More precisely, if one spin flips a subset of spins $s\subseteq\{1,\ldots,N\}$, then we consider $\ket{\Psi}\mapsto \sigma_y(s_1)\cdots\sigma_y(s_n)\ket{\Psi^*}$, and the corresponding $n$-tangle is
\begin{equation}
    \tau_n^{(s)} = \left|\Psi^\dagger\,\left[\bigotimes_{i=1}^n\sigma_y(s_i)\right]\,\Psi^*\right|^2.
\end{equation}
Moreover, one can study entanglement over size $n\leq N$ subsystems via an average over combinations, 
$\tau_n = \sum_s \tau_n^{(s)}$. 
}

\subsection{\it The single-angle limit: an integrable system}

{

In a non-homogeneous environment like a core-collapse supernova, the dependence of the pairwise $\nu$-$\nu$ interaction strength on their respective trajectories implies that the flavor evolution of a neutrino can, in general, depend on the emission angles (polar and azimuthal) with respect to the radial direction. This dependence vastly increases the number of degrees of freedom in the problem, and therefore, a common workaround---the \emph{single-angle approximation}---is to replace all of the trajectory-dependent interaction strengths among pairs of neutrinos with a single, appropriately chosen classical average. In this limit, one can define a trajectory-averaged interaction parameter $\bar\mu \equiv (\sqrt{2} G_F N/V) \langle 1 - \mathbf{\widehat p} \cdot \mathbf{\widehat q} \rangle$, which then depends only on the distance from the source. The Hamiltonian can then be re-written in a more simplified form:
\begin{equation} \label{eq:saham}
H = \sum_{\omega_\mathbf{p}} \omega_\mathbf{p} \, \vec{B}\cdot\vec{J}_{\omega_\mathbf{p}} + \frac{\bar\mu}{N} \, \vec{J} \cdot \vec{J}~.
\end{equation}
where $\vec{J} = \sum_{\omega_\mathbf{p}} \vec J_{\omega_\mathbf{p}}$ is the total neutrino isospin. Since the flavor evolution becomes trajectory-independent with this approximation, the neutrinos can be indexed simply by the \textit{magnitudes} of their momenta (or by their vacuum oscillation frequencies $\omega_\mathbf{p}$). For a neutrino source with spherically symmetric emission from a single surface of radius $R_\nu$ (a.k.a. the \lq\lq neutrino bulb\rq\rq---often used to model supernova neutrino emission~\cite{Duan:2006jv,Duan:2006b}), the trajectory averaging leads to the following expression for $\bar\mu$:
\begin{equation}
    \bar\mu(r) = \mu_0\left[1-\sqrt{1-\left(\frac{R_\nu}{r}\right)^2}\right]^2,
    \label{eq:samu}
\end{equation}
where $r$ is the distance from the center of the source. We also define $\mu_0\equiv (G_F/\sqrt{2})(N/V)=\bar\mu(R_\nu)$ to be the interaction strength at $r=R_\nu$. Here, we also assume time-invariance of the neutrino emission from the source; otherwise $\bar\mu$ would depend on both radius $r$ and time $t$. This invariance can be a reasonably good approximation since the timescales over which the emission changes significantly [$\mathcal{O}$(s)] are much longer than the light-crossing times across the supernova envelope [$\mathcal{O}(10$ ms)]. In the units of $\omega_0$, a typical scale for the vacuum oscillation frequency,\footnote{ For a $10$\,MeV neutrino energy and the atmospheric mass-squared splitting $\delta m^2 \simeq 2.3 \times 10^{-4}$\,eV$^2$, this scale is $\omega_0 \sim 10^{-16}$\,{MeV}.} $\mu_0$ can range from $\sim 10^6\omega_0$ during the neutronization burst phase of a core-collapse supernova to $\mu_0\sim10^5\mbox{--}10^4\omega_0$ during the late-time neutrino-driven wind phase.

The Hamiltonian from Eq.~\eqref{eq:saham} has been shown to possess a number of invariants (or conserved charges)~\cite{Pehlivan:2011}, analogous to the \lq\lq Gaudin magnets\rq\rq~\cite{Gaudin76} that had been previously identified as the conserved charges of the pairing-force Hamiltonian in nuclear and condensed-matter physics~\cite{Richardson63,Richardson64,Richardson65}. These conserved charges signify the \emph{integrability} of the Hamiltonian, which implies that exact eigenvalues and eigenstates may in principle be obtained in terms of closed-form solutions to a set of algebraic \lq\lq\emph{Bethe-Ansatz}\rq\rq\ equations~\cite{Bethe31}. In the context of a single-angle neutrino Hamiltonian, these procedures have been described and numerically implemented in recent literature~\cite{Pehlivan:2011,Birol:2018qhx,Patwardhan:2019zta}. These methods (and the associated parallels with other integrable many-body Hamiltonians in physics) have facilitated calculations of the neutrino flavor spectral split starting from an initial all-electron-flavor many-body state~\cite{Birol:2018qhx}, and yielded an explanation of this split as a Bardeen-Cooper-Schrieffer (BCS)-Bose-Einstein Condensate (BEC) crossover-type phenomenon~\cite{Pehlivan:2016lxx}.

In the single-angle limit, flavor evolution in small-sized systems [$\mathcal O(10\mbox{--}20)$ neutrinos, distributed across just as many vacuum oscillation modes] has been studied either using exact diagonalization methods~\cite{Patwardhan:2019zta,Cervia:2019}, or through brute-force numerical integration (e.g., fourth-order Runge-Kutta with an adaptive time step)~\cite{Rrapaj:2020,Patwardhan:2021rej}. The amount of entanglement between neutrinos in these calculations is seen to be correlated with the extent of the deviations from the mean-field behaviour of the system~\cite{Cervia:2019,Rrapaj:2020} and with the location of the spectral splits in the neutrino energy distributions~\cite{Patwardhan:2021rej}. An illustration of such a calculation with eight neutrinos distributed across uniformly-spaced oscillation frequencies, is depicted in Fig.~\ref{fig8nu}.

A particular class of models where the single-angle limit becomes exact is one where the neutrinos are grouped into two beams, and thus there is only one intersection-angle in the system. Since the neutrinos within each beam are assumed to always exist in a fully symmetrized state, the scaling of the number of independent amplitudes with neutrino number $N$ becomes a bit more favourable, i.e., $\sim (N/N_\text{beams})^{N_\text{beams} - 1}$ in the absence of vacuum mixing, or the same expression with $N_\text{beams} \rightarrow N_\text{beams} + 1$ if vacuum mixing is included~\cite{Xiong:2021evk}. This picture permits calculations with up to $\mathcal O(10^6)$ neutrinos in this setup, and, for different choices of initial conditions and/or neutrino mixing parameters, the many-body evolution for this configuration has been shown to either converge to or deviate from the mean-field limit~\cite{Xiong:2021evk,Martin:2021bri}.

}


\subsection{\it Multi-angle effects}
{
The assumption of uniform, trajectory-independent couplings induces additional symmetries in the system, otherwise not present. Such symmetries can partition the Hilbert space into disconnected sectors and limit neutrino flavor entanglement. When the assumption is lifted, the number of invariants of motion is greatly reduced relative to the single-angle approximation, further complicating the analysis. In such cases, one can treat exactly the time evolution only for systems of up to $N\sim20$ neutrinos~\cite{Rrapaj:2020} in full generality. Short of large scale quantum computers, capable of quickly exploring the exponentially large Hilbert space, one can focus on few beam systems. In such cases the number of neutrinos may increase, but the number of momenta directions is kept rather low. Even in these simplified setups, rather surprising results can be found, where the large-$N$ behavior is dependent on the angle between beams. In Ref.~\cite{Roggero:2022}, the authors found that in dense neutrino systems (ignoring vacuum oscillations), the beyond-mean-field behavior scales logarithmically with system  size, in sharp contradiction with a similar setup for bipolar oscillations~\cite{Friedland:2003eh}. This result is an indication of dynamical phase transitions and coincides with the instabilities in the linear analysis of the mean field equations, which we will describe below in more detail.}

{
More recently in Ref.~\cite{Martin:2023ljq}, the evolution of a $N = 16$ neutrino system with \textit{randomly chosen} one- and two-body couplings has been analyzed. For an initial condition comprised of some neutrinos as $\ket{\nu_e}$ and the rest as $\ket{\nu_x}$, the random coupling (i.e., multi-angle) result was shown to have a distinctly different asymptotic behaviour compared to its single-angle counterpart. This difference can be attributed to a dephasing effect in the mass basis, which is prevented in the single-angle case due to the integrability of the Hamiltonian.
}


\subsection{\it Flavor instabilities and dynamical phase transitions}

{
In the mean-field approach, collective neutrino oscillations are typically associated with unstable modes in the linear stability analysis of
the Hamiltonian described in Eq.~\eqref{eq:ham}.
 These instabilities are able to amplify initially small flavor perturbations exponentially quickly (e.g.,~\cite{Sawyer:2004,Sawyer:2005jk,Duan:2010,Chakraborty:2016yeg,Izaguirre:2017,Tamborra:2020cul,Richers:2022zug} and references therein). 
The presence of the forward-scattering interaction can allow collective effects to develop when $\mu \gtrsim \omega_\mathbf{p}$, giving rise to interesting phenomena like synchronization~\cite{Pastor:2002,Fuller:2006,Raffelt:2010,AKHMEDOV:2016}, bipolar oscillations~\cite{Kostelecky:1995,Duan:2006b,Duan:2007b}, and spectral splits/swaps~\cite{Duan:2006c,Duan:2007,Raffelt:2007,Dasgupta:2009,Martin:2020}.
}
On the other hand, in descriptions of interacting neutrino systems that permit many-body quantum dynamics, oscillations that develop on \lq\lq fast\rq\rq\ timescales are generally associated with rapid dynamical development of the neutrino entanglement entropy~\cite{Cervia:2019,Rrapaj:2020,Roggero:2021,Roggero:2021asb,Patwardhan:2021rej}. In Ref.~\cite{Roggero:2022}, rapid entanglement and mean field instabilities were also found to be linked for certain angular setups.

\begin{figure*}[htb]
    \centering
    \includegraphics[width=0.49\textwidth]{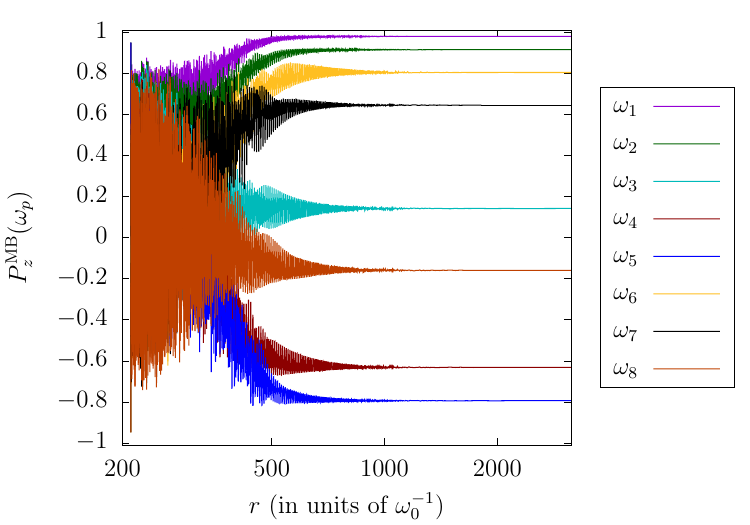}~
    \includegraphics[width=0.49\textwidth]{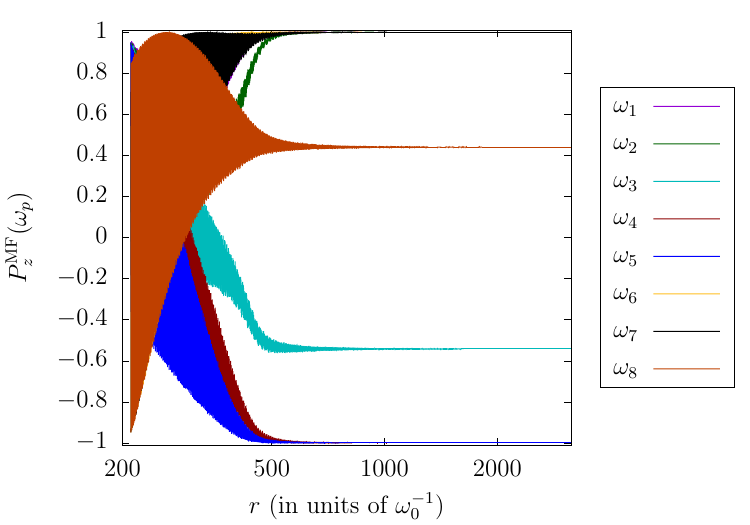}
    \includegraphics[width=0.49\textwidth]{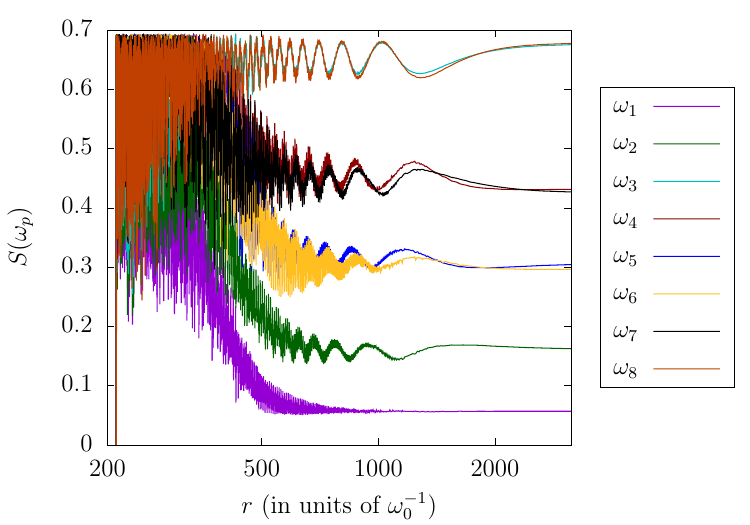}~
    \includegraphics[width=0.49\textwidth]{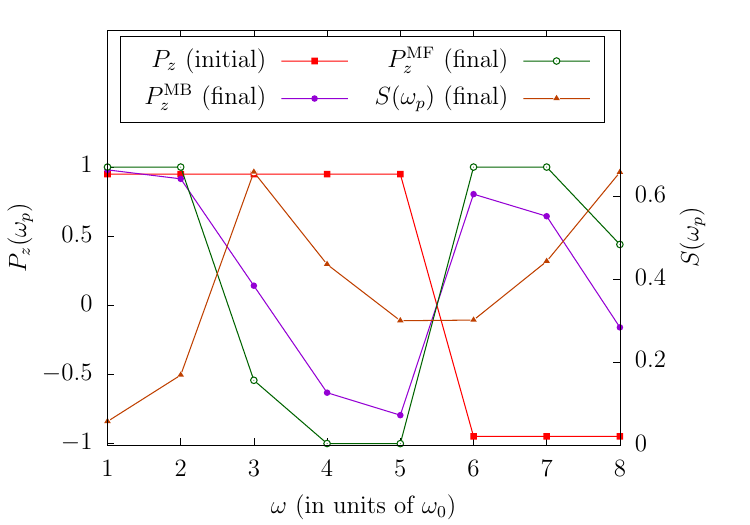}
    \caption{Evolution of an initial state 
    $\ket{\nu_e}^{\otimes6}\ket{\nu_x}^{\otimes2}$ 
    from a starting radius $r_0$ such that $\mu(r_0)=5\omega_0$, 
    with a small mixing angle ($\theta=0.161$) and 
    discrete, equally spaced oscillation frequencies $\omega_k = k\,\omega_0$, and a time-varying neutrino interaction strength $\mu(r)$ motivated by the neutrino bulb model~\cite{Duan:2006c}, in the single-angle approximation according to Eqs.~\eqref{eq:saham} and \eqref{eq:samu}. 
    Details of this calculation can be found in~\cite{Cervia:2019}. 
    \textbf{Top left:} Evolution of the $z$-components of the neutrino isospin expectation values (also known as \lq\lq Polarization vectors\rq\rq) in the mass basis, i.e., $P_z \equiv 2\langle J_z \rangle$, for the full many-body quantum system. \textbf{Top right:} Same as top left, but in the mean-field approximation. \textbf{Bottom left:} Evolution of the entanglement entropy of each neutrino, with respect to the rest of the ensemble. \textbf{Bottom right:} Asymptotic values of $P_z$ vs $\omega_k$, in the full many-body calculation (purple), and in the mean-field approximation (green), together with the initial $P_z$ values (red), and the asymptotic entanglement entropies (dark orange). Neutrinos located closest to the spectral splits in the energy distributions (in this case, at $\omega_2$ and $\omega_7$) develop the largest amount of entanglement and thereby experience the most significant deviations compared to their mean-field evolution.}
    \label{fig8nu}
\end{figure*}

{ Using a two-beam model, it was demonstrated \cite{Roggero:2021,Roggero:2021asb} that, when the frequency difference between two neutrino beams is comparable to the  $\nu$-$\nu$ interaction coupling, $\delta \omega \lesssim \mu$ to be precise, rapid and strong flavor oscillations develop for certain initial conditions. 
This finding was understood in terms of the presence of a Dynamic Phase Transition (DPT)~\cite{Heyl:2013,Heyl:2018}, which can be characterized by the introduction of the Loschmidt echo},
\begin{equation}
\label{eq:loch}
\mathcal{L}(t) = \left|\langle \Phi\lvert \exp\left(-\mathrm{i} t H\right)\rvert\Phi\rangle \right|^2\;,
\end{equation}
with $|\Phi\rangle$ the initial state at $t=0$. The quantity $\mathcal{L}(t)$ is a fidelity measure~\cite{Gorin:2006} that quantifies the probability for the system to return to its initial state. { For systems with degenerate initial state, as in both the two-beam bipolar case for $\delta \omega=0$~\cite{Roggero:2021,Roggero:2021asb} or the three-beam unstable case~\cite{Roggero:2022}, a suitable generalization of this quantity is obtained as follows (see Refs.~\cite{Heyl2014,Zunkovic2018,PhysRevD.104.123023}) 
\begin{equation}
\label{eq:echoes}
    \mathcal L_k(t) = \left|\langle\Phi_k | \Psi(t) \rangle \right|^2\;.
\end{equation}
where $\ket{\Phi_k}$ are the two degenerate states: one is the initial state $\ket{\Phi_0}=\ket{\Psi(0)}$, and the other one is orthogonal to $\ket{\Psi(0)}$. 
A DPT is then characterized by non-analyticities in the rate function
\begin{equation}
\label{eq:loch_rate}
\lambda(t) = -\frac{1}{N}\log\left[\mathcal{L}(t)\right]\;,
\end{equation}
where $N$ is the total number of particles in the system and $\lambda(t)$ is an intensive \lq\lq free energy\rq\rq~\cite{Heyl:2013,Gambassi:2012}. Here, the rate $\lambda(t)$ plays the role of a non-equilibrium equivalent of the thermodynamic free-energy. In the generalized case, $\lambda(t)$ is the minimum of the two options. Notably, other definitions of DPT are possible, such as time-averaged order parameters~\cite{Sciolla:2011,Sciolla:2013,Zunkovic:2018}. }

\subsection{\it Flavor evolution and entanglement in phase space}

{
 Recently, the neutrino flavor evolution and entanglement in this problem have also been analyzed using a \lq\lq phase space\rq\rq\ description~\cite{Lacroix:2022krq}, with the phase space coordinates being angles $\theta$ and $\phi$ that describe rotations in flavor space ($\theta = 0,\pi$ correspond to the basis states $\ket{\nu_{1,2}}$, respectively). For an interacting two-neutrino-beam setup, the Husimi quasi-probability or \lq\lq Q\rq\rq\ representation~\cite{Husimi:1940264} was constructed for the reduced density operator $\rho_A$ of neutrinos in one of the beams, as follows:
\begin{equation}
    Q_A(\Omega,t) = \bra{\Omega}\rho_A(t)\ket{\Omega},
\end{equation}
where $\ket{\Omega} = \ket{\theta,\phi}$ are coherent states, which form an over-complete basis with the closure relation
\begin{equation}
    \frac{2J_A + 1}{4\pi} \int d\Omega \ket{\Omega}\bra{\Omega} = 1.
\end{equation}
In the limit of infinite neutrino number, the Q representation reduces to a classical phase-space probability distribution. It was demonstrated that, for this system, the quasi-probability distribution remains relatively localized at early times, before subsequently de-localizing and developing a multi-modal structure with several peaks. This behavior is indicative of non-Gaussian entanglement, suggesting the presence of significant dynamics beyond the first and second moments of neutrino observables in the long-term evolution of this system. 

Based on these insights, an approximate method for estimating the long-term evolution of this system was also proposed. This method involves replacing the full quantum solution with a classical statistical average of several mean-field solutions, derived from a Gaussian distribution of initial conditions around the exact starting point of the system~\cite{Lacroix:2014sxa}. The time-evolution of one- and two-body observables obtained using this method was found to agree with the exact solution at early times, while also capturing in long-term evolution of these observables in a qualitative sense.
}

\section{Compact representations for many-body systems}


{ To calculate quantum corrections beyond the mean-field coherent limit, 
one can systematically incorporate $n$-body density matrices $\rho_{1\ldots n}$ for $n \geq 1$, given by
\begin{equation}
    \rho_{1\ldots n} = \frac{N!}{(N-n)!}\mathrm{Tr}_{n+1\ldots N}\rho_{1\ldots N},
\end{equation}
into the coupled equations of motion for $N$ neutrinos, as follows~\cite{Volpe:2013uxl}:
\begin{equation}
    \mathrm{i}\partial_t\rho_{1\ldots n} = [H_{1\ldots n},\rho_{1\ldots n}]+\sum_{s=1}^n\mathrm{Tr}_{n+1}[V(s,n+1),\rho_{1\ldots n+1}],
\end{equation}
where $H_{1\ldots n}$ is the Hamiltonian truncated for the first $n$ neutrinos in a given ordering and $V(i,j)$ is the two-body interaction potential for a pair of neutrinos $(i,j)$. 
This procedure follows the Bogoliubov-Born-Green-Kirkwood-Yvon (BBGKY) hierarchy for density matrices: the mean-field equations can be recovered by restricting to $n=2$ and approximating $\rho_{12}\approx\rho_1\rho_2$ 
(i.e., requiring the two-body correlation function to be zero). }
{ 
In this description, investigating the importance of quantum corrections would involve incorporating the $n$-body density operators for progressively increasing values of $n$, while checking for convergence of results for physical observables. }

\begin{figure*}
    \centering
    \includegraphics[width=0.985\textwidth]{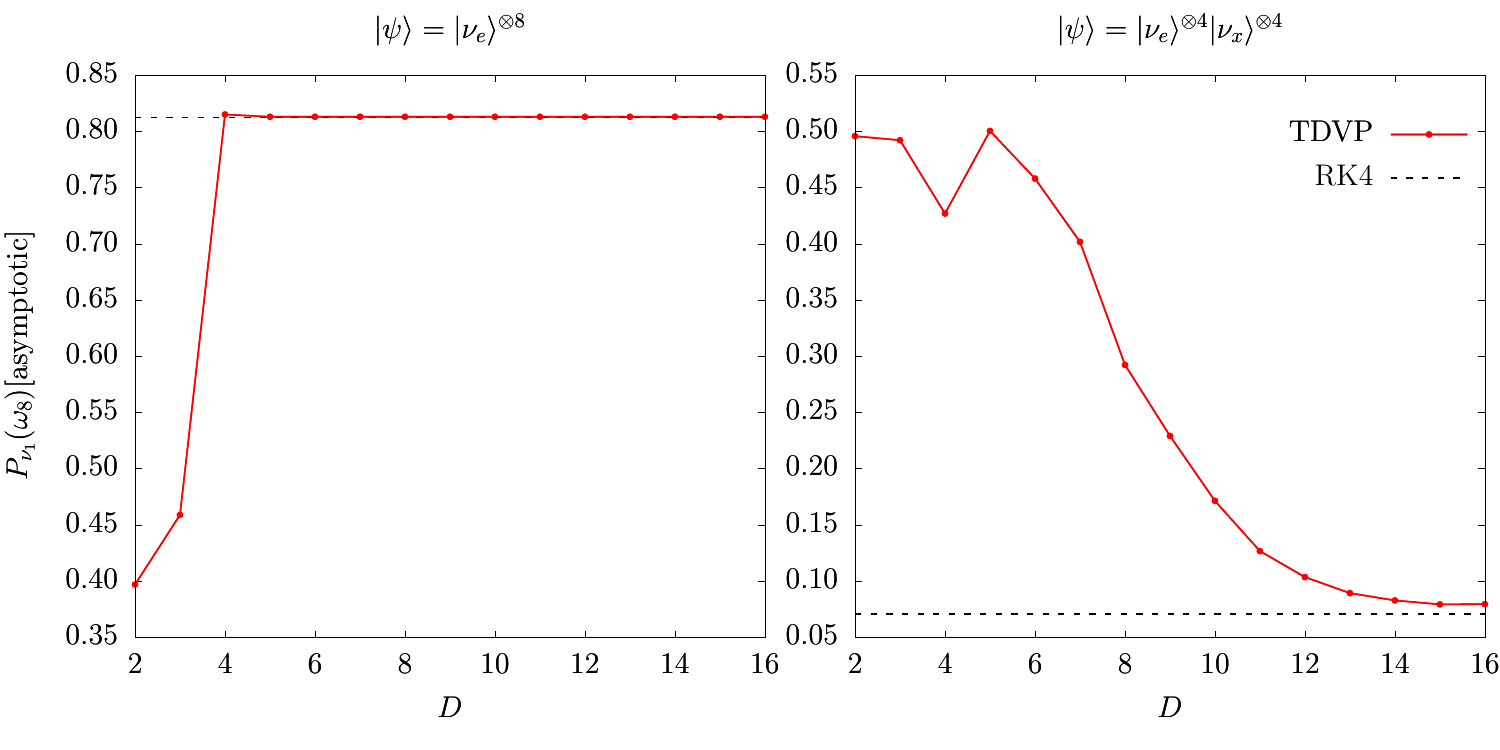}
    \caption{The survival probability in first mass eigenstate in a eight neutrino system as a function of bond dimension $(D)$ with an initial state $\ket{\nu_{e}}^{\otimes 8}$ (left) and $\ket{\nu_{e}}^{\otimes 4}\ket{\nu_{x}}^{\otimes 4}$ (right) with time step $\delta t=0.1$. The results are compared with the ones obtained from RK4.}
    \label{fig:N8itensor}
\end{figure*}
{ Because of the exponential growth ($2^n$) of the size of the Hilbert space, classical computers are unable to exactly simulate systems of more than $\sim 30$ neutrinos. One possibility is to introduce compact representations of the wave function through tensor network methods~\cite{Roggero:2021,Roggero:2021asb,Cervia:2022}, and more specifically matrix product states~\cite{Vidal:2003,2011AnPhy.326...96S,PAECKEL2019167998}. These methods allow for the computation of systems of hundreds of neutrinos in a two-beam setup with a time-independent $\nu$-$\nu$ interaction~\cite{Roggero:2021,Roggero:2021asb}. Alternatively, when considering very dense neutrino gases (if vacuum oscillations are ignored), methods based on generalized angular momentum representations, by analogy between two flavor oscillations and spin systems (see~Eq.\ref{eq:psi_angular_momentum_basis}), can reach up to $\mathcal O(10^6)$ neutrinos and predict the thermodynamic limit in some cases~\cite{Friedland:2003eh,Friedland:2006ke,Xiong:2021evk,Martin:2021bri,Roggero:2022}.}     

{In the case of a \textit{time-dependent} interaction strength, 
a more sophisticated tensor network method, namely, the time-dependent variational principle (TDVP) method has been utilized in Ref.~\cite{Cervia:2022}. These techniques provided considerable computational benefit for an initial state with all neutrinos in the same flavor, allowing for evolution of a system with $\approx50$ oscillation modes. This limit was a consequence of the entanglement among neutrinos being more localized in certain regions of the neutrino energy distribution. For systems with initial states being a mixture of $\nu_e$ and $\nu_x$ flavors, the entanglement is more de-localized, and therefore the comparative advantage gained through TDVP methods is less dramatic, although work remains in progress on this front.}  
{ As shown in Fig.~\ref{fig:N8itensor}, for a system of eight neutrinos with an initial state $\ket{\psi}=\ket{\nu_{e}}^{\otimes 8}$ the survival probability for the neutrino in the highest frequency mode in the first mass eigenstate $P_{\nu_{1}}$, converges to the exact value obtained from RK4 for a very small bond dimension $D\sim 4$. Therefore, $D$ can be truncated to a significantly small number (for $N=8$ the maximum bond dimension is $2^4=16$) in the case of an initial state with all neutrinos in the same flavor. In the case of an initial state with half neutrinos in $\nu_{e}$ and $\nu_{x}$ flavor each, the $P_{\nu_{1}}$ does not converge to the exact value until $D\sim15$. Furthermore, the results do not converge to those of RK4 exactly, and hence the truncation requires considering a smaller time step for more accurate results.} 

{ Non-vanishing connected correlations, being the essential characteristic of beyond mean-field behavior, deserve particular attention. Ursell functions~\cite{Shlosman1986} provide a unified framework for $n$-connected correlations, 
 \begin{equation}
 \begin{split}
  C_n(\sigma_z)=\frac{\partial^n}{\partial \lambda_1 \ldots\partial \lambda_n} \text{ln} \left(\langle \Psi | e^{\sum_{i=1}^n\lambda_i {\sigma_z}_i} | \Psi \rangle \right) \bigg|_{\bold{\lambda}=\bold{0}}, \
  \sigma_z = 2 J_z
 \end{split}
\end{equation} and have been studied in recent works~\cite{Illa:2022zgu}. It is worth pointing out that, despite only pairwise interactions in the system, higher order correlations dynamically develop as well. 
}

\section{ Quantum simulations of many-body neutrino systems} ~\label{sec:qsim}
{
For generic closed quantum systems, quantum simulation algorithms are promising tools to study quantum many-body evolution. 
Preliminary attempts~\cite{Hall:2021rbv,Yeter-Aydeniz:2021olz,Illa:2022jqb,Amitrano:2022yyn,Illa:2022zgu} to simulate collective neutrino oscillations on a quantum computer have already been taken, for small system sizes and short evolution times. 
In Ref.~\cite{Hall:2021rbv}, a system of four neutrinos was simulated using superconducting qubit hardware. 
In particular, the unitary evolution operator $U(t)=\exp(-\mathrm{i}Ht)$ was approximated via first-order Trotter-Suzuki decomposition, with error of $\mathcal{O}(t^{2})$. 

Notably, since the interaction in this model is long-range, quantum devices with all-to-all connectivity are desirable. 
Nevertheless, on a quantum device having connectivity only among neighboring qubits, SWAP operations can still be used to implement this interaction~\cite{Hall:2021rbv}, though doing so requires more quantum gates in the simulation that may decohere the quantum state being simulated. 
}
{
Alternatively, hybrid quantum algorithms such as  quantum Lanczos (QLanczos) could be used~\cite{Yeter-Aydeniz:2021olz} to approximately diagonalize the neutrino many-body Hamiltonian on a quantum computer. 
Further, real-time evolution using trotterization may allow for calculations of transition probabilities for interacting neutrinos. 
However, practical limitations of current quantum hardware prevent studies of larger systems in these earlier quantum simulations, i.e., limited number of unitary operations with low accuracy. 
}
{ More recently in Refs.~\cite{Amitrano:2022yyn} and \cite{Illa:2022zgu}, trapped-ion quantum devices were utilized to perform the simulations eight and twelve neutrinos respectively, thanks to the all-to-all qubit connectivity in trapped-ion based architecture.}

\section{ Three flavor case} \label{sec:3fla}
\begin{figure*}[t]
    \centering
    \includegraphics[width=0.985\columnwidth]{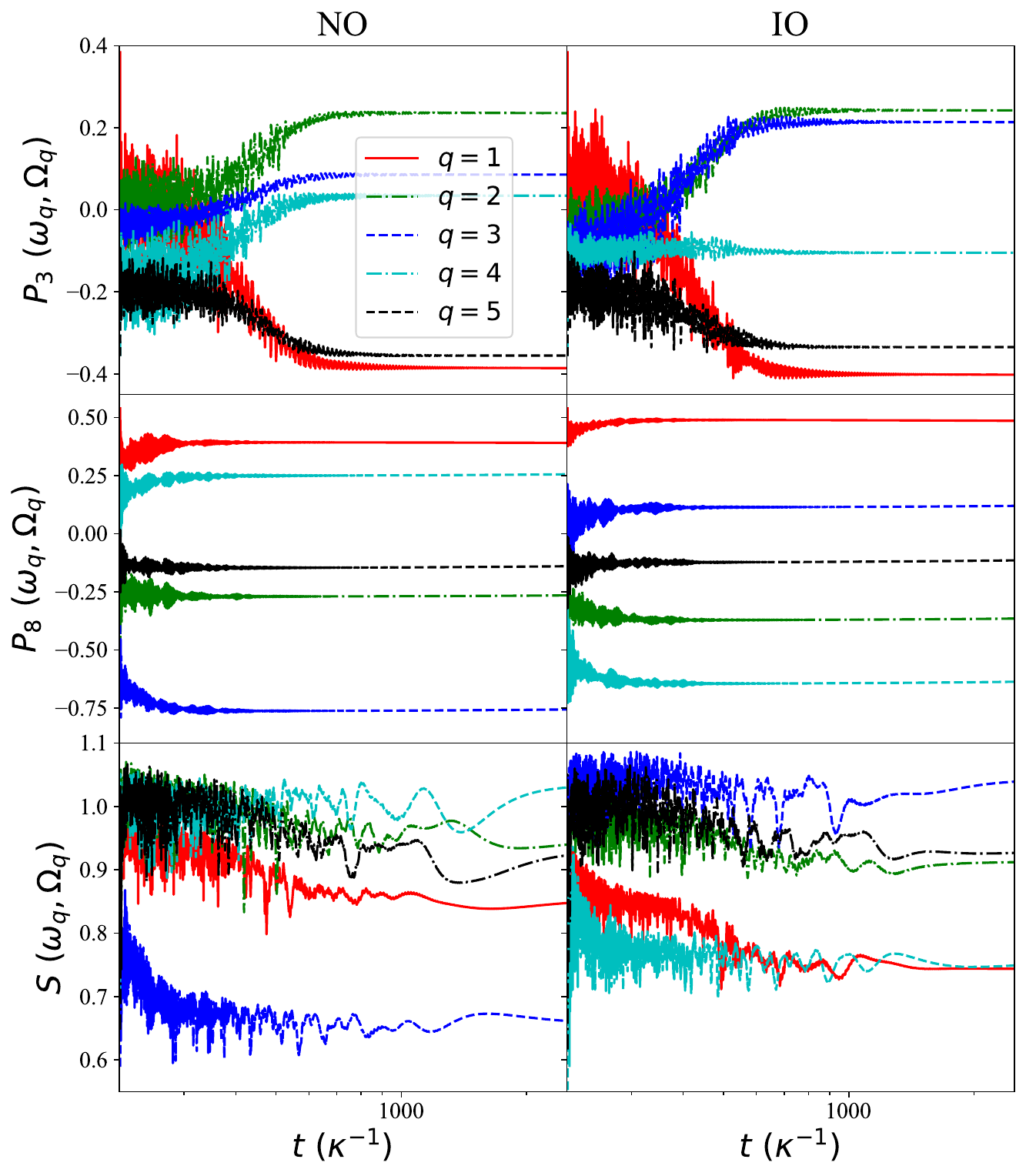}
    \includegraphics[width=0.97\columnwidth]{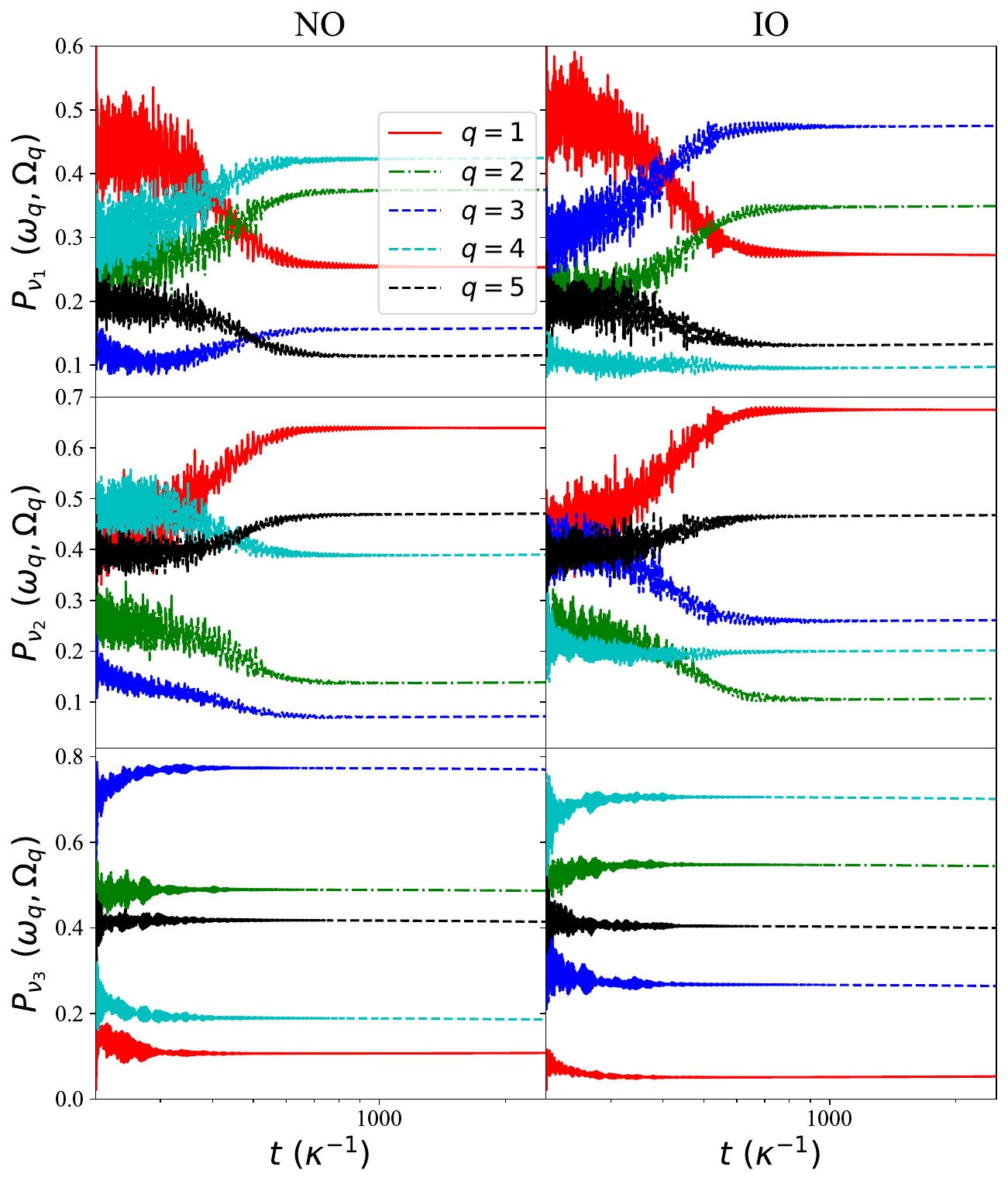}
    \caption{The temporal evolution of $P_{3}$ (top), $P_{8}$ (middle) and $S$ (bottom) for $N=5$ neutrinos with initial state $\ket{\psi}=\ket{\nu_{e}\nu_{\mu}\nu_{\mu}\nu_{\tau}\nu_{\tau}}$ in NO (left) and IO (right). {\it Right panels:} The temporal evolution $P_{\nu_{1}}$ (top), $P_{\nu_{2}}$ (middle) and $P_{\nu_{3}}$ (bottom) for $N=5$ neutrinos with initial state $\ket{\psi}=\ket{\nu_{e}\nu_{\mu}\nu_{\mu}\nu_{\tau}\nu_{\tau}}$ in NO (left) and IO (right).}
    \label{fig:N5_mix}
\end{figure*}

{ An extension of the frequently adopted two-flavor framework to three flavors in the mean-field approximation has revealed several unique phenomena, e.g., multiple spectral splits, in collective neutrino oscillations~\cite{Fogli:2008fj, Duan:2008prl, Dasgupta:2008prd, Dasgupta:2009prl, Dasgupta:2010prd, Friedland:2010prl, Airen:2018nvp, Chakraborty:2019wxe, Shalgar:2021wlj}. To see whether these differences translate to the many-body picture, the neutrino many-body problem has recently been analyzed in a three-flavor setting~\cite{Siwach:prd2023}. The Hamiltonian given in Eq.~\eqref{eq:saham} can be generalized to the three-flavor case:
\begin{equation}\label{eq:H3}
    H = \sum_{p}\Vec{B}\cdot\Vec{Q_{p}}+\sum_{p,p'}\mu(r)\Vec{Q}\cdot\Vec{Q} \ ,
\end{equation}
where the generators $Q_{i'}$ are given by
\begin{equation}
    Q_{i'}=\frac{1}{2}\sum_{i,j=1}^{3}a^{\dagger}_{i}(\lambda_{i'})_{ij}a_{j} \ ,
\end{equation}
with $\lambda$'s as the Gell-Mann matrices.
The auxiliary vector $\vec{B}$ is given by
\begin{equation}\label{eq:B3}
    \Vec{B}=\left(0,0,\omega_{p},0,0,0,0,\Omega_{p}\right) \ .
\end{equation}
Here, the oscillation frequencies are
\begin{subequations}
\label{eq:constants} 
\begin{eqnarray}
    \omega_{p}&=&-\frac{1}{2E} \delta m^{2} \ , \label{eq:wp} \\    
    \Omega_{p}&=&-\frac{1}{2E}\Delta m^{2} \label{eq:Wp} \ ,
\end{eqnarray}
\end{subequations}
where $\delta m^{2} = m_{2}^{2} - m_{1}^{2}$, and $\Delta m^{2} \approx |m_{3}^{2}-m_{2}^{2}| \approx |m_{3}^{2}-m_{1}^{2}|$, and $E$ is the energy of neutrino. 

Due to computational limitations, only a system of up to five neutrinos could be considered in this treatment. To quantify the entanglement, the von-Neumann entanglement entropies $S(\omega_q)$ and the components of the neutrino polarization vectors $\vec P_q$ are calculated (straightforward generalizations of the definitions in Sec.~\ref{sec:quant}) . In the three-flavor case, two components of the \textit{total} polarization vector (of the entire ensemble), namely $P_{3}$ and $P_{8}$, are conserved through the course of flavor evolution, whereas, in the two-flavor case only the total $P_{z}$ is conserved. It was found that the entanglement in the three-flavor case can be significantly larger in comparison to the two-flavor case for an initial condition with all neutrinos in the electron flavor. 

Here we further investigate the entanglement in three-flavor neutrino many-body system with an initial state different from those considered in Ref.~\cite{Siwach:prd2023}.
Shown in Fig.~\ref{fig:N5_mix} are the results for the time-evolution of a five-neutrino system with an initial state $\ket{\psi}=\ket{\nu_{e}\nu_{\mu}\nu_{\mu}\nu_{\tau}\nu_{\tau}}$ in both normal (NO) and inverted (IO) mass orderings. { The vacuum oscillation frequencies of the neutrinos are chosen to be $\omega_q = q\,\omega_p$ and $\Omega_q = q\,\Omega_p$, with $\omega_p$ and $\Omega_p$ defined by Eq.~\eqref{eq:constants}, with a neutrino energy $E=10\,$MeV. $\kappa = 10^{-17}\,$MeV is a suitably chosen scaling factor in terms of which all the other dimensionful quantities are defined in the numerical calculations.} The ordering of the asymptotic $P_{3}$ values with respect to the frequency modes $q$ is observed to be the same in both NO and IO. However, the magnitudes are significantly different. The $P_{8}$ values on the other hand show differences in both magnitude and ordering. One can see from the entanglement entropies $S(\omega_q)$ that the neutrino in frequency mode $q=4$ (3) is maximally entangled in NO (IO), whereas this neutrino is the \textit{least} entangled in IO (NO), respectively. Further information on the entanglement can be obtained from the graphical representation of projection of polarization vectors $P_{3}$ and $P_{8}$ values in the $\hat{e}_{3}$-$\hat{e}_{8}$ plane, as given in Ref.~\cite{Siwach:prd2023}. 

From the probabilities of finding a neutrino in a particular mass eigenstate, $P_{\nu_{i}}$, as shown in the right panels of Fig.~\ref{fig:N5_mix}, the mixing of different mass eigenstates can be investigated. In NO, the neutrino which is maximally entangled, {\it i.e.} $q=4$, has $\sim 40\%$ contribution from first and second mass eigenstates each and $\sim 20\%$ contribution from third mass eigenstate. The least entangled $q=4$ neutrino has $\sim 80\%$ contribution from third mass eigenstate and $\sim 10\%$ contribution from first and second mass eigenstates each. In IO, the maximally entangled neutrino in $q=3$ frequency mode has $\sim 50\%$ contribution from the first mass eigenstate and $\sim 25\%$ contribution from second and third mass eigenstates each. Therefore, the results in different mass orderings differ significantly. 

We notice that the entanglement in many-body neutrino systems depends substantially on the initial state. These differences enhance in three-flavor case as compared to the two-flavor case. For example, the entanglement entropy for $N=5$ neutrino system with an initial state $\ket{\nu_{e}\nu_{e}\nu_{\mu}\nu_{\mu}\nu_{\tau}}$ (see Ref.~\cite{Siwach:prd2023}) in the case of NO is maximum for the neutrino in the highest frequency mode. Whereas in the case of an initial state $\ket{\nu_{e}\nu_{\mu}\nu_{\mu}\nu_{\tau}\nu_{\tau}}$, the entropy is maximum for the second highest frequency mode $q=4$ neutrino. Several other differences can be observed in other properties and in inverted mass ordering. Therefore, to better understand the neutrino many-body system, similar calculations should be performed for a significantly larger system considering various initial states.
}

\section{Concluding remarks} \label{sec:concl}

{ Many-body quantum dynamics of dense neutrino systems is an active area of research that in recent years has shown a rapid development in terms of understanding and analyzing many body correlations, instabilities, and dynamical phase transitions, with various groups attempting to investigate the problem using different types of classical and quantum computational tools. Methods adapted from quantum information have shown great potential for further studying many body effects.} { The ongoing quest to augment our understanding of these quantum dynamics, and collective flavor oscillations in general, stems from the importance of these phenomena in influencing the neutrino transport and nucleosynthesis in astrophysical environments like core-collapse supernovae and binary neutron star mergers, which are important open problems in the domain of high-energy/nuclear astrophysics.}


\begin{acknowledgements}
This work was supported in part by the U.S. Department of Energy, Office of Science, Office of High Energy Physics, under Award No. DE-SC0019465. AVP acknowledges support from the U.S. Department of Energy under contract number DE-AC02-76SF00515. MJC: the U.S. Department of Energy, Office of Nuclear Physics under Award Number DE-SC0021143. ABB acknowledges support from the NSF grants PHY-2108339 and PHY-2020275. ER work was funded by iTHEMS RIKEN.
\end{acknowledgements}




\bibliographystyle{spphys}
\bibliography{references}

\begin{thebibliography}{100}
\providecommand{\url}[1]{{#1}}
\providecommand{\urlprefix}{URL }
\expandafter\ifx\csname urlstyle\endcsname\relax
  \providecommand{\doi}[1]{DOI \discretionary{}{}{}#1}\else
  \providecommand{\doi}{DOI \discretionary{}{}{}\begingroup
  \urlstyle{rm}\Url}\fi

\bibitem{Janka:2006fh}
H.T. Janka, K.~Langanke, A.~Marek, G.~Martinez-Pinedo, B.~Mueller, Phys. Rept.
  \textbf{442}, 38 (2007).
\newblock \doi{10.1016/j.physrep.2007.02.002}.
\newblock astro-ph/0612072

\bibitem{Burrows:2020qrp}
A.~Burrows, D.~Vartanyan, Nature \textbf{589}(7840), 29 (2021).
\newblock \doi{10.1038/s41586-020-03059-w}.
\newblock 2009.14157

\bibitem{Fuller:2022nbn}
G.M. Fuller, W.C. Haxton,   (2022).
\newblock 2208.08050

\bibitem{Foucart:2022bth}
F.~Foucart, Living Reviews in Computational Astrophysics \textbf{9}(1) (2023).
\newblock \doi{10.1007/s41115-023-00016-y}.
\newblock \urlprefix\url{http://dx.doi.org/10.1007/s41115-023-00016-y}.
\newblock 2209.02538

\bibitem{Kyutoku:2017voj}
K.~Kyutoku, K.~Kiuchi, Y.~Sekiguchi, M.~Shibata, K.~Taniguchi, Phys. Rev. D
  \textbf{97}(2), 023009 (2018).
\newblock \doi{10.1103/PhysRevD.97.023009}.
\newblock 1710.00827

\bibitem{Grohs:2015tfy}
E.~Grohs, G.M. Fuller, C.T. Kishimoto, M.W. Paris, A.~Vlasenko, Phys. Rev. D
  \textbf{93}(8), 083522 (2016).
\newblock \doi{10.1103/PhysRevD.93.083522}.
\newblock 1512.02205

\bibitem{Surman:2003qt}
R.~Surman, G.C. McLaughlin, Astrophys. J. \textbf{603}, 611 (2004).
\newblock \doi{10.1086/381672}.
\newblock astro-ph/0308004

\bibitem{Martinez-Pinedo:2017ksl}
G.~Mart\'\i{}nez-Pinedo, T.~Fischer, K.~Langanke, A.~Lohs, A.~Sieverding, M.R.
  Wu, \emph{{Neutrinos and Their Impact on Core-Collapse Supernova
  Nucleosynthesis}} (2017).
\newblock \doi{10.1007/978-3-319-21846-5_78}

\bibitem{Kajino:2014bra}
T.~Kajino, G.J. Mathews, T.~Hayakawa, J. Phys. G \textbf{41}, 044007 (2014).
\newblock \doi{10.1088/0954-3899/41/4/044007}

\bibitem{Frohlich:2015spx}
C.~Fr\"ohlich, J.~Casanova, M.~Hempel, M.~Liebend\"orfer, C.A. Melton,
  A.~Perego, AIP Conf. Proc. \textbf{1604}(1), 178 (2015).
\newblock \doi{10.1063/1.4883428}

\bibitem{Langanke:2019ggn}
K.~Langanke, G.~Martinez-Pinedo, A.~Sieverding,   (2019).
\newblock \doi{10.22661/AAPPSBL.2018.28.6.41}.
\newblock 1901.03741

\bibitem{Roberts:2016igt}
L.F. Roberts, J.~Lippuner, M.D. Duez, J.A. Faber, F.~Foucart, J.C. Lombardi,
  S.~Ning, C.D. Ott, M.~Ponce, Mon. Not. Roy. Astron. Soc. \textbf{464}(4),
  3907 (2017).
\newblock \doi{10.1093/mnras/stw2622}.
\newblock 1601.07942

\bibitem{Steigman:2012ve}
G.~Steigman, Adv. High Energy Phys. \textbf{2012}, 268321 (2012).
\newblock \doi{10.1155/2012/268321}.
\newblock 1208.0032

\bibitem{Qian:1993dg}
Y.Z. Qian, G.M. Fuller, G.J. Mathews, R.~Mayle, J.R. Wilson, S.E. Woosley,
  Phys. Rev. Lett. \textbf{71}, 1965 (1993).
\newblock \doi{10.1103/PhysRevLett.71.1965}

\bibitem{Yoshida:2006qz}
T.~Yoshida, T.~Kajino, H.~Yokomakura, K.~Kimura, A.~Takamura, D.H. Hartmann,
  Phys. Rev. Lett. \textbf{96}, 091101 (2006).
\newblock \doi{10.1103/PhysRevLett.96.091101}.
\newblock astro-ph/0602195

\bibitem{Duan:2010af}
H.~Duan, A.~Friedland, G.~McLaughlin, R.~Surman, J. Phys. G \textbf{38}, 035201
  (2011).
\newblock \doi{10.1088/0954-3899/38/3/035201}.
\newblock 1012.0532

\bibitem{Kajino:2012zz}
T.~Kajino, W.~Aoki, M.K. Cheoun, S.~Chiba, W.~Fujiya, T.~Hayakawa, G.J.
  Mathews, K.~Nakamura, K.~Shaku, T.~Yoshida, AIP Conf. Proc. \textbf{1441}(1),
  375 (2012).
\newblock \doi{10.1063/1.3700559}

\bibitem{Wu:2014kaa}
M.R. Wu, Y.Z. Qian, G.~Martinez-Pinedo, T.~Fischer, L.~Huther, Phys. Rev. D
  \textbf{91}(6), 065016 (2015).
\newblock \doi{10.1103/PhysRevD.91.065016}.
\newblock 1412.8587

\bibitem{Sasaki:2017jry}
H.~Sasaki, T.~Kajino, T.~Takiwaki, T.~Hayakawa, A.B. Balantekin, Y.~Pehlivan,
  Phys. Rev. D \textbf{96}(4), 043013 (2017).
\newblock \doi{10.1103/PhysRevD.96.043013}.
\newblock 1707.09111

\bibitem{Balantekin:2017bau}
A.B. Balantekin, AIP Conf. Proc. \textbf{1947}(1), 020012 (2018).
\newblock \doi{10.1063/1.5030816}.
\newblock 1710.04108

\bibitem{Xiong:2019nvw}
Z.~Xiong, M.R. Wu, Y.Z. Qian,   (2019).
\newblock \doi{10.3847/1538-4357/ab2870}.
\newblock 1904.09371

\bibitem{Xiong:2020ntn}
Z.~Xiong, A.~Sieverding, M.~Sen, Y.Z. Qian, Astrophys. J. \textbf{900}(2), 144
  (2020).
\newblock \doi{10.3847/1538-4357/abac5e}.
\newblock 2006.11414

\bibitem{George:2020veu}
M.~George, M.R. Wu, I.~Tamborra, R.~Ardevol-Pulpillo, H.T. Janka, Phys. Rev. D
  \textbf{102}(10), 103015 (2020).
\newblock \doi{10.1103/PhysRevD.102.103015}.
\newblock 2009.04046

\bibitem{Wolfenstein:1977ue}
L.~Wolfenstein, Phys. Rev. D \textbf{17}, 2369 (1978).
\newblock \doi{10.1103/PhysRevD.17.2369}

\bibitem{Mikheyev:1985zog}
S.P. Mikheyev, A.Y. Smirnov, Sov. J. Nucl. Phys. \textbf{42}, 913 (1985)

\bibitem{Mikheev:1986if}
S.P. Mikheev, A.Y. Smirnov, Sov. Phys. JETP \textbf{64}, 4 (1986).
\newblock 0706.0454

\bibitem{Notzold:1987ik}
D.~Notzold, G.~Raffelt, Nucl. Phys. B \textbf{307}, 924 (1988).
\newblock \doi{10.1016/0550-3213(88)90113-7}

\bibitem{Fuller:!987aa}
G.M. {Fuller}, R.W. {Mayle}, J.R. {Wilson}, D.N. {Schramm}, Astrophys. J.
  \textbf{322}, 795 (1987).
\newblock \doi{10.1086/165772}

\bibitem{Pantaleone:1992xh}
J.T. Pantaleone, Phys. Rev. D \textbf{46}, 510 (1992).
\newblock \doi{10.1103/PhysRevD.46.510}

\bibitem{Pantaleone:1992eq}
J.T. Pantaleone, Phys. Lett. B \textbf{287}, 128 (1992).
\newblock \doi{10.1016/0370-2693(92)91887-F}

\bibitem{Samuel:1993}
S.~Samuel, Phys. Rev. D \textbf{48}, 1462 (1993).
\newblock \doi{10.1103/PhysRevD.48.1462}.
\newblock \urlprefix\url{https://link.aps.org/doi/10.1103/PhysRevD.48.1462}

\bibitem{Duan:2009cd}
H.~Duan, J.P. Kneller, J. Phys. G \textbf{36}, 113201 (2009).
\newblock \doi{10.1088/0954-3899/36/11/113201}.
\newblock 0904.0974

\bibitem{Duan:2010}
H.~Duan, G.M. Fuller, Y.Z. Qian, Annual Review of Nuclear and Particle Science
  \textbf{60}(1), 569 (2010).
\newblock \doi{10.1146/annurev.nucl.012809.104524}.
\newblock \urlprefix\url{https://doi.org/10.1146/annurev.nucl.012809.104524}.
\newblock https://doi.org/10.1146/annurev.nucl.012809.104524

\bibitem{Chakraborty:2016yeg}
S.~Chakraborty, R.~Hansen, I.~Izaguirre, G.~Raffelt, Nucl. Phys. B
  \textbf{908}, 366 (2016).
\newblock \doi{10.1016/j.nuclphysb.2016.02.012}.
\newblock 1602.02766

\bibitem{Tamborra:2020cul}
I.~Tamborra, S.~Shalgar, Ann. Rev. Nucl. Part. Sci. \textbf{71}, 165 (2021).
\newblock \doi{10.1146/annurev-nucl-102920-050505}.
\newblock 2011.01948

\bibitem{Richers:2022zug}
S.~Richers, M.~Sen,  pp. 1--17 (2022).
\newblock \doi{10.1007/978-981-15-8818-1_125-1}.
\newblock \urlprefix\url{https://doi.org/10.1007/978-981-15-8818-1_125-1}.
\newblock 2207.03561

\bibitem{Bauer:2022hpo}
C.W. Bauer, et~al.,   (2022).
\newblock 2204.03381

\bibitem{Sigl:1993ctk}
G.~Sigl, G.~Raffelt, Nucl. Phys. B \textbf{406}, 423 (1993).
\newblock \doi{10.1016/0550-3213(93)90175-O}

\bibitem{Qian:1994wh}
Y.Z. Qian, G.M. Fuller, Phys. Rev. D \textbf{51}, 1479 (1995).
\newblock \doi{10.1103/PhysRevD.51.1479}.
\newblock astro-ph/9406073

\bibitem{Shalgar:2023ooi}
S.~Shalgar, I.~Tamborra,   (2023).
\newblock 2304.13050

\bibitem{Balantekin_2006}
A.B. Balantekin, Y.~Pehlivan, Journal of Physics G: Nuclear and Particle
  Physics \textbf{34}(1), 47 (2006).
\newblock \doi{10.1088/0954-3899/34/1/004}.
\newblock \urlprefix\url{https://doi.org/10.1088/0954-3899/34/1/004}

\bibitem{Johns:2023ewj}
L.~Johns,   (2023).
\newblock 2305.04916

\bibitem{Bell:2003mg}
N.F. Bell, A.A. Rawlinson, R.F. Sawyer, Phys. Lett. B \textbf{573}, 86 (2003).
\newblock \doi{10.1016/j.physletb.2003.08.035}.
\newblock hep-ph/0304082

\bibitem{Friedland:2003dv}
A.~Friedland, C.~Lunardini, Phys. Rev. D \textbf{68}, 013007 (2003).
\newblock \doi{10.1103/PhysRevD.68.013007}.
\newblock hep-ph/0304055

\bibitem{Friedland:2003eh}
A.~Friedland, C.~Lunardini, JHEP \textbf{10}, 043 (2003).
\newblock \doi{10.1088/1126-6708/2003/10/043}.
\newblock hep-ph/0307140

\bibitem{Friedland:2006ke}
A.~Friedland, B.H.J. McKellar, I.~Okuniewicz, Phys. Rev. D \textbf{73}, 093002
  (2006).
\newblock \doi{10.1103/PhysRevD.73.093002}.
\newblock hep-ph/0602016

\bibitem{McKellar:2009py}
B.H.J. McKellar, I.~Okuniewicz, J.~Quach, Phys. Rev. D \textbf{80}, 013011
  (2009).
\newblock \doi{10.1103/PhysRevD.80.013011}.
\newblock 0903.3139

\bibitem{Duan:2006jv}
H.~Duan, G.M. Fuller, J.~Carlson, Y.Z. Qian, Phys. Rev. Lett. \textbf{97},
  241101 (2006).
\newblock \doi{10.1103/PhysRevLett.97.241101}.
\newblock astro-ph/0608050

\bibitem{Duan:2006c}
H.~Duan, G.M. Fuller, J.~Carlson, Y.Z. Qian, Phys. Rev. D \textbf{74}, 105014
  (2006).
\newblock \doi{10.1103/PhysRevD.74.105014}.
\newblock \urlprefix\url{https://link.aps.org/doi/10.1103/PhysRevD.74.105014}

\bibitem{Duan:2007}
H.~Duan, G.M. Fuller, J.~Carlson, Y.Z. Qian, Phys. Rev. Lett. \textbf{99},
  241802 (2007).
\newblock \doi{10.1103/PhysRevLett.99.241802}.
\newblock
  \urlprefix\url{https://link.aps.org/doi/10.1103/PhysRevLett.99.241802}

\bibitem{Raffelt:2007}
G.G. Raffelt, A.Y. Smirnov, Phys. Rev. D \textbf{76}, 081301 (2007).
\newblock \doi{10.1103/PhysRevD.76.081301}.
\newblock \urlprefix\url{https://link.aps.org/doi/10.1103/PhysRevD.76.081301}

\bibitem{Raffelt:2007xt}
G.G. Raffelt, A.Y. Smirnov, Phys. Rev. D \textbf{76}, 125008 (2007).
\newblock \doi{10.1103/PhysRevD.76.125008}.
\newblock 0709.4641

\bibitem{Sawyer:2004}
R.F. Sawyer,   (2004).
\newblock hep-ph/0408265

\bibitem{Eisert:2015}
J.~Eisert, M.~Friesdorf, C.~Gogolin, Nature Physics \textbf{11}(2), 124 (2015).
\newblock \doi{10.1038/nphys3215}.
\newblock \urlprefix\url{https://doi.org/10.1038/nphys3215}

\bibitem{Cervia:2019}
M.J. Cervia, A.V. Patwardhan, A.B. Balantekin, S.N. Coppersmith, C.W. Johnson,
  Phys. Rev. D \textbf{100}, 083001 (2019).
\newblock \doi{10.1103/PhysRevD.100.083001}.
\newblock \urlprefix\url{https://link.aps.org/doi/10.1103/PhysRevD.100.083001}

\bibitem{Rrapaj:2020}
E.~Rrapaj, Phys. Rev. C \textbf{101}, 065805 (2020).
\newblock \doi{10.1103/PhysRevC.101.065805}.
\newblock \urlprefix\url{https://link.aps.org/doi/10.1103/PhysRevC.101.065805}

\bibitem{Roggero:2021asb}
A.~Roggero, Phys. Rev. D \textbf{104}(10), 103016 (2021).
\newblock \doi{10.1103/PhysRevD.104.103016}.
\newblock 2102.10188

\bibitem{Roggero:2022}
A.~Roggero, E.~Rrapaj, Z.~Xiong, Phys. Rev. D \textbf{106}, 043022 (2022).
\newblock \doi{10.1103/PhysRevD.106.043022}.
\newblock \urlprefix\url{https://link.aps.org/doi/10.1103/PhysRevD.106.043022}

\bibitem{Nielsen:2011:QCQ:1972505}
M.A. Nielsen, I.L. Chuang, \emph{Quantum Computation and Quantum Information:
  10th Anniversary Edition} (Cambridge University Press, New York, 2011)

\bibitem{2011AnPhy.326...96S}
U.~{Schollw{\"o}ck}, Ann.~Phys. (Amsterdam) \textbf{326}(1), 96 (2011).
\newblock \doi{10.1016/j.aop.2010.09.012}

\bibitem{PAECKEL2019167998}
S.~Paeckel, T.~K\"ohler, A.~Swoboda, S.R. Manmana, U.~Schollw\"ock, C.~Hubig,
  Ann.~Phys.~(N.~Y.) \textbf{411}, 167998 (2019).
\newblock \doi{https://doi.org/10.1016/j.aop.2019.167998}.
\newblock
  \urlprefix\url{https://www.sciencedirect.com/science/article/pii/S0003491619302532}

\bibitem{Vidal:2003lvx}
G.~Vidal, Phys. Rev. Lett. \textbf{93}, 040502 (2004).
\newblock \doi{10.1103/PhysRevLett.93.040502}.
\newblock quant-ph/0310089

\bibitem{Schuch:2008zza}
N.~Schuch, M.M. Wolf, F.~Verstraete, J.I. Cirac, Phys. Rev. Lett. \textbf{100},
  030504 (2008).
\newblock \doi{10.1103/PhysRevLett.100.030504}.
\newblock 0705.0292

\bibitem{PhysRevA.63.044301}
A.~Wong, N.~Christensen, Phys. Rev. A \textbf{63}, 044301 (2001).
\newblock \doi{10.1103/PhysRevA.63.044301}.
\newblock \urlprefix\url{https://link.aps.org/doi/10.1103/PhysRevA.63.044301}

\bibitem{Illa:2022zgu}
M.~Illa, M.J. Savage, Phys. Rev. Lett. \textbf{130}(22), 221003 (2023).
\newblock \doi{10.1103/PhysRevLett.130.221003}.
\newblock 2210.08656

\bibitem{Duan:2006b}
H.~Duan, G.M. Fuller, Y.Z. Qian, Phys. Rev. D \textbf{74}, 123004 (2006).
\newblock \doi{10.1103/PhysRevD.74.123004}.
\newblock \urlprefix\url{https://link.aps.org/doi/10.1103/PhysRevD.74.123004}

\bibitem{Pehlivan:2011}
Y.~Pehlivan, A.B. Balantekin, T.~Kajino, T.~Yoshida, Phys. Rev. D \textbf{84},
  065008 (2011).
\newblock \doi{10.1103/PhysRevD.84.065008}.
\newblock \urlprefix\url{https://link.aps.org/doi/10.1103/PhysRevD.84.065008}

\bibitem{Gaudin76}
{Gaudin, M.}, J. Phys. France \textbf{37}(10), 1087 (1976).
\newblock \doi{10.1051/jphys:0197600370100108700}.
\newblock \urlprefix\url{https://doi.org/10.1051/jphys:0197600370100108700}

\bibitem{Richardson63}
R.W. {Richardson}, Physics Letters \textbf{3}(6), 277 (1963).
\newblock \doi{10.1016/0031-9163(63)90259-2}

\bibitem{Richardson64}
R.W. {Richardson}, N.~{Sherman}, Nuclear Physics \textbf{52}, 221 (1964).
\newblock \doi{10.1016/0029-5582(64)90687-X}

\bibitem{Richardson65}
R.W. {Richardson}, Journal of Mathematical Physics \textbf{6}(7), 1034 (1965).
\newblock \doi{10.1063/1.1704367}

\bibitem{Bethe31}
H.~{Bethe}, Zeitschrift fur Physik \textbf{71}(3-4), 205 (1931).
\newblock \doi{10.1007/BF01341708}

\bibitem{Birol:2018qhx}
S.~Birol, Y.~Pehlivan, A.B. Balantekin, T.~Kajino, Phys. Rev. D \textbf{98}(8),
  083002 (2018).
\newblock \doi{10.1103/PhysRevD.98.083002}.
\newblock 1805.11767

\bibitem{Patwardhan:2019zta}
A.V. Patwardhan, M.J. Cervia, A.~Baha~Balantekin, Phys. Rev. D \textbf{99}(12),
  123013 (2019).
\newblock \doi{10.1103/PhysRevD.99.123013}.
\newblock 1905.04386

\bibitem{Pehlivan:2016lxx}
Y.~Pehlivan, A.L. Suba\c{s}\i{}, N.~Ghazanfari, S.~Birol, H.~Y\"uksel, Phys.
  Rev. D \textbf{95}(6), 063022 (2017).
\newblock \doi{10.1103/PhysRevD.95.063022}.
\newblock 1603.06360

\bibitem{Patwardhan:2021rej}
A.V. Patwardhan, M.J. Cervia, A.B. Balantekin, Phys. Rev. D \textbf{104}(12),
  123035 (2021).
\newblock \doi{10.1103/PhysRevD.104.123035}.
\newblock 2109.08995

\bibitem{Xiong:2021evk}
Z.~Xiong, Phys. Rev. D \textbf{105}(10), 103002 (2022).
\newblock \doi{10.1103/PhysRevD.105.103002}.
\newblock 2111.00437

\bibitem{Martin:2021bri}
J.D. Martin, A.~Roggero, H.~Duan, J.~Carlson, V.~Cirigliano, Phys. Rev. D
  \textbf{105}(8), 083020 (2022).
\newblock \doi{10.1103/PhysRevD.105.083020}.
\newblock 2112.12686

\bibitem{Martin:2023ljq}
J.D. Martin, A.~Roggero, H.~Duan, J.~Carlson,   (2023).
\newblock 2301.07049

\bibitem{Sawyer:2005jk}
R.F. Sawyer, Phys. Rev. D \textbf{72}, 045003 (2005).
\newblock \doi{10.1103/PhysRevD.72.045003}.
\newblock hep-ph/0503013

\bibitem{Izaguirre:2017}
I.~Izaguirre, G.~Raffelt, I.~Tamborra, Phys. Rev. Lett. \textbf{118}, 021101
  (2017).
\newblock \doi{10.1103/PhysRevLett.118.021101}.
\newblock
  \urlprefix\url{https://link.aps.org/doi/10.1103/PhysRevLett.118.021101}

\bibitem{Pastor:2002}
S.~Pastor, G.~Raffelt, D.V. Semikoz, Phys. Rev. D \textbf{65}, 053011 (2002).
\newblock \doi{10.1103/PhysRevD.65.053011}.
\newblock \urlprefix\url{https://link.aps.org/doi/10.1103/PhysRevD.65.053011}

\bibitem{Fuller:2006}
G.M. Fuller, Y.Z. Qian, Phys. Rev. D \textbf{73}, 023004 (2006).
\newblock \doi{10.1103/PhysRevD.73.023004}.
\newblock \urlprefix\url{https://link.aps.org/doi/10.1103/PhysRevD.73.023004}

\bibitem{Raffelt:2010}
G.G. Raffelt, I.~Tamborra, Phys. Rev. D \textbf{82}, 125004 (2010).
\newblock \doi{10.1103/PhysRevD.82.125004}.
\newblock \urlprefix\url{https://link.aps.org/doi/10.1103/PhysRevD.82.125004}

\bibitem{AKHMEDOV:2016}
E.~Akhmedov, A.~Mirizzi, Nuclear Physics B \textbf{908}, 382 (2016).
\newblock \doi{https://doi.org/10.1016/j.nuclphysb.2016.02.011}.
\newblock
  \urlprefix\url{https://www.sciencedirect.com/science/article/pii/S0550321316000559}.
\newblock Neutrino Oscillations: Celebrating the Nobel Prize in Physics 2015

\bibitem{Kostelecky:1995}
V.A. Kosteleck\'y, S.~Samuel, Phys. Rev. D \textbf{52}, 621 (1995).
\newblock \doi{10.1103/PhysRevD.52.621}.
\newblock \urlprefix\url{https://link.aps.org/doi/10.1103/PhysRevD.52.621}

\bibitem{Duan:2007b}
H.~Duan, G.M. Fuller, J.~Carlson, Y.Z. Qian, Phys. Rev. D \textbf{75}, 125005
  (2007).
\newblock \doi{10.1103/PhysRevD.75.125005}.
\newblock \urlprefix\url{https://link.aps.org/doi/10.1103/PhysRevD.75.125005}

\bibitem{Dasgupta:2009}
B.~Dasgupta, A.~Dighe, G.G. Raffelt, A.Y. Smirnov, Phys. Rev. Lett.
  \textbf{103}, 051105 (2009).
\newblock \doi{10.1103/PhysRevLett.103.051105}.
\newblock
  \urlprefix\url{https://link.aps.org/doi/10.1103/PhysRevLett.103.051105}

\bibitem{Martin:2020}
J.D. Martin, J.~Carlson, H.~Duan, Phys. Rev. D \textbf{101}, 023007 (2020).
\newblock \doi{10.1103/PhysRevD.101.023007}.
\newblock \urlprefix\url{https://link.aps.org/doi/10.1103/PhysRevD.101.023007}

\bibitem{Roggero:2021}
A.~Roggero, Phys. Rev. D \textbf{104}, 123023 (2021).
\newblock \doi{10.1103/PhysRevD.104.123023}.
\newblock \urlprefix\url{https://link.aps.org/doi/10.1103/PhysRevD.104.123023}

\bibitem{Heyl:2013}
M.~Heyl, A.~Polkovnikov, S.~Kehrein, Phys. Rev. Lett. \textbf{110}, 135704
  (2013).
\newblock \doi{10.1103/PhysRevLett.110.135704}.
\newblock
  \urlprefix\url{https://link.aps.org/doi/10.1103/PhysRevLett.110.135704}

\bibitem{Heyl:2018}
M.~Heyl, Reports on Progress in Physics \textbf{81}(5), 054001 (2018).
\newblock \doi{10.1088/1361-6633/aaaf9a}.
\newblock \urlprefix\url{https://doi.org/10.1088/1361-6633/aaaf9a}

\bibitem{Gorin:2006}
T.~Gorin, T.~Prosen, T.H. Seligman, M.~Žnidarič, Physics Reports
  \textbf{435}(2), 33 (2006).
\newblock \doi{https://doi.org/10.1016/j.physrep.2006.09.003}.
\newblock
  \urlprefix\url{https://www.sciencedirect.com/science/article/pii/S0370157306003310}

\bibitem{Heyl2014}
M.~Heyl, Phys. Rev. Lett. \textbf{113}, 205701 (2014).
\newblock \doi{10.1103/PhysRevLett.113.205701}.
\newblock
  \urlprefix\url{https://link.aps.org/doi/10.1103/PhysRevLett.113.205701}

\bibitem{Zunkovic2018}
B.~\ifmmode \check{Z}\else \v{Z}\fi{}unkovi\ifmmode~\check{c}\else \v{c}\fi{},
  M.~Heyl, M.~Knap, A.~Silva, Phys. Rev. Lett. \textbf{120}, 130601 (2018).
\newblock \doi{10.1103/PhysRevLett.120.130601}.
\newblock
  \urlprefix\url{https://link.aps.org/doi/10.1103/PhysRevLett.120.130601}

\bibitem{PhysRevD.104.123023}
A.~Roggero, Phys. Rev. D \textbf{104}, 123023 (2021).
\newblock \doi{10.1103/PhysRevD.104.123023}.
\newblock \urlprefix\url{https://link.aps.org/doi/10.1103/PhysRevD.104.123023}

\bibitem{Gambassi:2012}
A.~Gambassi, A.~Silva, Phys. Rev. Lett. \textbf{109}, 250602 (2012).
\newblock \doi{10.1103/PhysRevLett.109.250602}.
\newblock
  \urlprefix\url{https://link.aps.org/doi/10.1103/PhysRevLett.109.250602}

\bibitem{Sciolla:2011}
B.~Sciolla, G.~Biroli, Journal of Statistical Mechanics: Theory and Experiment
  \textbf{2011}(11), P11003 (2011).
\newblock \doi{10.1088/1742-5468/2011/11/p11003}.
\newblock \urlprefix\url{https://doi.org/10.1088/1742-5468/2011/11/p11003}

\bibitem{Sciolla:2013}
B.~Sciolla, G.~Biroli, Phys. Rev. B \textbf{88}, 201110 (2013).
\newblock \doi{10.1103/PhysRevB.88.201110}.
\newblock \urlprefix\url{https://link.aps.org/doi/10.1103/PhysRevB.88.201110}

\bibitem{Zunkovic:2018}
B.~\ifmmode \check{Z}\else \v{Z}\fi{}unkovi\ifmmode~\check{c}\else \v{c}\fi{},
  M.~Heyl, M.~Knap, A.~Silva, Phys. Rev. Lett. \textbf{120}, 130601 (2018).
\newblock \doi{10.1103/PhysRevLett.120.130601}.
\newblock
  \urlprefix\url{https://link.aps.org/doi/10.1103/PhysRevLett.120.130601}

\bibitem{Lacroix:2022krq}
D.~Lacroix, A.B. Balantekin, M.J. Cervia, A.V. Patwardhan, P.~Siwach, Phys.
  Rev. D \textbf{106}(12), 123006 (2022).
\newblock \doi{10.1103/PhysRevD.106.123006}.
\newblock 2205.09384

\bibitem{Husimi:1940264}
K.~Husimi, Proceedings of the Physico-Mathematical Society of Japan. 3rd Series
  \textbf{22}(4), 264 (1940).
\newblock \doi{10.11429/ppmsj1919.22.4_264}

\bibitem{Lacroix:2014sxa}
D.~Lacroix, S.~Ayik, Eur. Phys. J. A \textbf{50}, 95 (2014).
\newblock \doi{10.1140/epja/i2014-14095-8}.
\newblock 1402.2393

\bibitem{Volpe:2013uxl}
C.~Volpe, D.~V\"a\"an\"anen, C.~Espinoza, Phys. Rev. D \textbf{87}(11), 113010
  (2013).
\newblock \doi{10.1103/PhysRevD.87.113010}.
\newblock 1302.2374

\bibitem{Cervia:2022}
M.J. Cervia, P.~Siwach, A.V. Patwardhan, A.B. Balantekin, S.N. Coppersmith,
  C.W. Johnson, Phys. Rev. D \textbf{105}, 123025 (2022).
\newblock \doi{10.1103/PhysRevD.105.123025}.
\newblock \urlprefix\url{https://link.aps.org/doi/10.1103/PhysRevD.105.123025}

\bibitem{Vidal:2003}
G.~Vidal, Phys. Rev. Lett. \textbf{91}, 147902 (2003).
\newblock \doi{10.1103/PhysRevLett.91.147902}.
\newblock
  \urlprefix\url{https://link.aps.org/doi/10.1103/PhysRevLett.91.147902}

\bibitem{Shlosman1986}
S.B. Shlosman, Communications in Mathematical Physics \textbf{102}(4), 679
  (1986).
\newblock \doi{10.1007/BF01221652}.
\newblock \urlprefix\url{https://doi.org/10.1007/BF01221652}

\bibitem{Hall:2021rbv}
B.~Hall, A.~Roggero, A.~Baroni, J.~Carlson, Phys. Rev. D \textbf{104}, 063009
  (2021).
\newblock \doi{10.1103/PhysRevD.104.063009}.
\newblock \urlprefix\url{https://link.aps.org/doi/10.1103/PhysRevD.104.063009}

\bibitem{Yeter-Aydeniz:2021olz}
K.~Yeter-Aydeniz, S.~Bangar, G.~Siopsis, R.C. Pooser, Quant. Inf. Proc.
  \textbf{21}(3), 84 (2022).
\newblock \doi{10.1007/s11128-021-03348-x}.
\newblock \urlprefix\url{https://doi.org/10.1007/s11128-021-03348-x}.
\newblock 2104.03273

\bibitem{Illa:2022jqb}
M.~Illa, M.J. Savage, Phys. Rev. A \textbf{106}, 052605 (2022).
\newblock \doi{10.1103/PhysRevA.106.052605}.
\newblock \urlprefix\url{https://link.aps.org/doi/10.1103/PhysRevA.106.052605}

\bibitem{Amitrano:2022yyn}
V.~Amitrano, A.~Roggero, P.~Luchi, F.~Turro, L.~Vespucci, F.~Pederiva, Phys.
  Rev. D \textbf{107}, 023007 (2023).
\newblock \doi{10.1103/PhysRevD.107.023007}.
\newblock \urlprefix\url{https://link.aps.org/doi/10.1103/PhysRevD.107.023007}

\bibitem{Fogli:2008fj}
G.~Fogli, E.~Lisi, A.~Marrone, I.~Tamborra, JCAP \textbf{04}, 030 (2009).
\newblock \doi{10.1088/1475-7516/2009/04/030}.
\newblock 0812.3031

\bibitem{Duan:2008prl}
H.~Duan, G.M. Fuller, J.~Carlson, Y.Z. Qian, Phys. Rev. Lett. \textbf{100},
  021101 (2008).
\newblock \doi{10.1103/PhysRevLett.100.021101}.
\newblock
  \urlprefix\url{https://link.aps.org/doi/10.1103/PhysRevLett.100.021101}

\bibitem{Dasgupta:2008prd}
B.~Dasgupta, A.~Dighe, A.~Mirizzi, G.G. Raffelt, Phys. Rev. D \textbf{77},
  113007 (2008).
\newblock \doi{10.1103/PhysRevD.77.113007}.
\newblock \urlprefix\url{https://link.aps.org/doi/10.1103/PhysRevD.77.113007}

\bibitem{Dasgupta:2009prl}
B.~Dasgupta, A.~Dighe, G.G. Raffelt, A.Y. Smirnov, Phys. Rev. Lett.
  \textbf{103}, 051105 (2009).
\newblock \doi{10.1103/PhysRevLett.103.051105}.
\newblock
  \urlprefix\url{https://link.aps.org/doi/10.1103/PhysRevLett.103.051105}

\bibitem{Dasgupta:2010prd}
B.~Dasgupta, A.~Mirizzi, I.~Tamborra, R.~Tom\`as, Phys. Rev. D \textbf{81},
  093008 (2010).
\newblock \doi{10.1103/PhysRevD.81.093008}.
\newblock \urlprefix\url{https://link.aps.org/doi/10.1103/PhysRevD.81.093008}

\bibitem{Friedland:2010prl}
A.~Friedland, Phys. Rev. Lett. \textbf{104}, 191102 (2010).
\newblock \doi{10.1103/PhysRevLett.104.191102}.
\newblock
  \urlprefix\url{https://link.aps.org/doi/10.1103/PhysRevLett.104.191102}

\bibitem{Airen:2018nvp}
S.~Airen, F.~Capozzi, S.~Chakraborty, B.~Dasgupta, G.~Raffelt, T.~Stirner, JCAP
  \textbf{12}, 019 (2018).
\newblock \doi{10.1088/1475-7516/2018/12/019}.
\newblock 1809.09137

\bibitem{Chakraborty:2019wxe}
M.~Chakraborty, S.~Chakraborty, JCAP \textbf{01}, 005 (2020).
\newblock \doi{10.1088/1475-7516/2020/01/005}.
\newblock 1909.10420

\bibitem{Shalgar:2021wlj}
S.~Shalgar, I.~Tamborra, Phys. Rev. D \textbf{104}(2), 023011 (2021).
\newblock \doi{10.1103/PhysRevD.104.023011}.
\newblock 2103.12743

\bibitem{Siwach:prd2023}
P.~Siwach, A.M. Suliga, A.B. Balantekin, Phys. Rev. D \textbf{107}, 023019
  (2023).
\newblock \doi{10.1103/PhysRevD.107.023019}.
\newblock \urlprefix\url{https://link.aps.org/doi/10.1103/PhysRevD.107.023019}

\end{thebibliography}

\end{document}